\PassOptionsToPackage{unicode}{hyperref}
\PassOptionsToPackage{hyphens}{url}
\PassOptionsToPackage{dvipsnames,svgnames,x11names}{xcolor}
\documentclass[12pt]{article}
\usepackage{setspace}
\usepackage{amsmath,amssymb,amsthm}  
\usepackage{bbm}
\usepackage{subcaption}
\usepackage{bm}
\usepackage{makecell}       
\usepackage{multirow}       
\usepackage{diagbox}        
\usepackage{tabularx}       
\usepackage{threeparttable} 
\usepackage[linesnumbered,ruled,vlined]{algorithm2e}
\usepackage{iftex}
\ifPDFTeX
  \usepackage[T1]{fontenc}
  \usepackage[utf8]{inputenc}
  \usepackage{textcomp} 
\else 
  \usepackage{unicode-math}
  \defaultfontfeatures{Scale=MatchLowercase}
  \defaultfontfeatures[\rmfamily]{Ligatures=TeX,Scale=1}
\fi
\usepackage{lmodern}
\ifPDFTeX\else  
\fi
\IfFileExists{upquote.sty}{\usepackage{upquote}}{}
\IfFileExists{microtype.sty}{
  \usepackage[]{microtype}
  \UseMicrotypeSet[protrusion]{basicmath} 
}{}
\makeatletter
\@ifundefined{KOMAClassName}{
  \IfFileExists{parskip.sty}{%
  }{
    \setlength{\parindent}{2em} 
    \setlength{\parskip}{6pt plus 2pt minus 1pt}}
}{
  \KOMAoptions{parskip=half}}
\makeatother
\usepackage{xcolor}
\makeatletter
\ifx\paragraph\undefined\else
  \let\oldparagraph\paragraph
  \renewcommand{\paragraph}{
    \@ifstar
      \xxxParagraphStar
      \xxxParagraphNoStar
  }
  \newcommand{\xxxParagraphStar}[1]{\oldparagraph*{#1}\mbox{}}
  \newcommand{\xxxParagraphNoStar}[1]{\oldparagraph{#1}\mbox{}}
\fi
\ifx\subparagraph\undefined\else
  \let\oldsubparagraph\subparagraph
  \renewcommand{\subparagraph}{
    \@ifstar
      \xxxSubParagraphStar
      \xxxSubParagraphNoStar
  }
  \newcommand{\xxxSubParagraphStar}[1]{\oldsubparagraph*{#1}\mbox{}}
  \newcommand{\xxxSubParagraphNoStar}[1]{\oldsubparagraph{#1}\mbox{}}
\fi
\makeatother

\usepackage{longtable,booktabs,array}
\usepackage{calc} 
\usepackage{etoolbox}
\makeatletter
\patchcmd\longtable{\par}{\if@noskipsec\mbox{}\fi\par}{}{}
\makeatother
\IfFileExists{footnotehyper.sty}{\usepackage{footnotehyper}}{\usepackage{footnote}}
\makesavenoteenv{longtable}
\usepackage{graphicx}
\makeatletter
\def\maxwidth{\ifdim\Gin@nat@width>\linewidth\linewidth\else\Gin@nat@width\fi}
\def\maxheight{\ifdim\Gin@nat@height>\textheight\textheight\else\Gin@nat@height\fi}
\makeatother
\setkeys{Gin}{width=\maxwidth,height=\maxheight,keepaspectratio}
\makeatletter
\def\fps@figure{htbp}
\makeatother

\makeatletter
\@ifpackageloaded{caption}{}{\usepackage{caption}}
\AtBeginDocument{%
\ifdefined\contentsname
  \renewcommand*\contentsname{Table of contents}
\else
  \newcommand\contentsname{Table of contents}
\fi
\ifdefined\listfigurename
  \renewcommand*\listfigurename{List of Figures}
\else
  \newcommand\listfigurename{List of Figures}
\fi
\ifdefined\listtablename
  \renewcommand*\listtablename{List of Tables}
\else
  \newcommand\listtablename{List of Tables}
\fi
\ifdefined\figurename
  \renewcommand*\figurename{Figure}
\else
  \newcommand\figurename{Figure}
\fi
\ifdefined\tablename
  \renewcommand*\tablename{Table}
\else
  \newcommand\tablename{Table}
\fi
}
\@ifpackageloaded{float}{}{\usepackage{float}}
\floatstyle{ruled}
\@ifundefined{c@chapter}{\newfloat{codelisting}{h}{lop}}{\newfloat{codelisting}{h}{lop}[chapter]}
\floatname{codelisting}{Listing}

\makeatother
\makeatletter
\@ifpackageloaded{caption}{}{\usepackage{caption}}
\@ifpackageloaded{subcaption}{}{\usepackage{subcaption}}
\makeatother

\ifLuaTeX
  \usepackage{selnolig}  
\fi
\usepackage[]{natbib}
\usepackage{xr-hyper}
\usepackage{bookmark}

\IfFileExists{xurl.sty}{\usepackage{xurl}}{} 
\hypersetup{
  pdftitle={Multi-Source Prediction-Powered Inference},
  pdfauthor={Wenhui Li; Fen Jiang; Xinyu Zhang},
  pdfkeywords={Inference; Model averaging; Machine learning},
  colorlinks=true,
  linkcolor={blue},
  filecolor={Maroon},
  citecolor={Blue},
  urlcolor={Blue},
  pdfcreator={LaTeX via pandoc}}

\newcommand{\anon}{1} 

\usepackage{makecell}

\newtheorem{assumption}{Assumption}

\newtheorem{theorem}{Theorem}

\newtheorem{example}{Example}

\newtheorem{corollary}{Corollary}

\newcommand{\bb}{\mathbf{b}}
\newcommand{\be}{\mathbf{e}}

\newcommand{\bh}{\mathbf{h}}

\newcommand{\bu}{\mathbf{u}}
\newcommand{\bv}{\mathbf{v}}
\newcommand{\bw}{\mathbf{w}}
\newcommand{\bx}{\mathbf{x}}

\newcommand{\by}{\mathbf{y}}
\newcommand{\bz}{\mathbf{z}}

\newcommand{\bA}{\mathbf{A}}

\newcommand{\bC}{\mathbf{C}}

\newcommand{\bI}{\mathbf{I}}

\newcommand{\bQ}{\mathbf{Q}}

\newcommand{\bX}{\mathbf{X}}

\newcommand{\bzero}{\mathbf{0}}

\newcommand{\bbeta}{\bm{\beta}}

\newcommand{\btheta}{\bm{\theta}}

\newcommand{\bmu}{\bm{\mu}}

\newcommand{\bSig}{\bm{\Sigma}}

\newcommand{\mE}{\mathbb{E}}

\newcommand{\mcP}{\mathcal{P}}
\newcommand{\mcQ}{\mathcal{Q}}
\newcommand{\mcN}{\mathcal{N}}
\newcommand{\mcD}{\mathcal{D}}
\newcommand{\mcI}{\mathcal{I}}
\newcommand{\mcX}{\mathcal{X}}

\newcommand{\mcW}{\mathcal{W}}
\newcommand{\mcC}{\mathcal{C}} 
\newcommand{\mcZ}{\mathcal{Z}}

\newcommand{\fs}{{f}^{(s)}}

\newcommand{\hTsk}{\widehat{T}^{(s),-[k]}}
\newcommand{\hSsk}{\widehat{S}^{(s),-[k]}}
\newcommand{\hRsk}{\widehat{\mathrm{DR}}^{(s),-[k]}}

\newcommand{\hbw}{\widehat \bw}
\newcommand{\hbQ}{\widehat \bQ}

\newcommand{\hbtheta}{\widehat \btheta}
\newcommand{\hbSig}{\widehat \bSig}

\newcommand{\var}{\mathrm{var}}
\newcommand{\cov}{\mathrm{cov}}

\newcommand{\diag}{\mathrm{diag}}

\DeclareMathOperator*{\argmin}{arg\,min}

\newcommand{\xis}{x_i^{(s)}}
\newcommand{\yis}{y_i^{(s)}}
\newcommand{\xizero}{\ensuremath{x_i^{(0)}}}
\newcommand{\yizero}{\ensuremath{y_i^{(0)}}}
\newcommand{\Ss}{S^{(s)}}
\newcommand{\Ts}{T^{(s)}}

\definecolor{mygreen}{RGB}{0,140,100}



\begin{document}

\def\spacingset#1{\renewcommand{\baselinestretch}%
{#1}\small\normalsize} \spacingset{1}


\if1\anon
{
  \title{\bf Multi-Source Prediction-Powered Inference}
  \author{\parbox{0.95\textwidth}{\centering\small
   Wenhui Li$^{a,b}$, Fen Jiang$^{c}$, and Xinyu Zhang$^{a,b,c\ast}$\\
   $^a$Center for Forecasting Science, \\Academy of Mathematics and Systems Science, Chinese Academy of Sciences;\\
   $^b$State Key Laboratory of Mathematical Sciences,\\ Academy of Mathematics and Systems Science, Chinese Academy of Sciences;\\ 
   $^c$School of Management,\\ University of Science and Technology of China}} 
  \maketitle
} \fi

\if0\anon
{
  \bigskip
  \bigskip
  \bigskip
  \begin{center}
    {\LARGE\bf Multi-Source Prediction-Powered Inference}
\end{center}
  \medskip
} \fi

\bigskip
\begin{abstract}
Prediction-powered inference integrates a small gold-standard dataset with large pseudo-labeled data, whose labels are generated by machine learning methods, to enhance statistical inference. In modern applications, multiple data sources and diverse machine learning methods often give rise to multiple pseudo-labeled datasets, each encoding potentially different aspects of the underlying information. However, how to optimally combine multiple data sources and machine learning methods for statistical inference remains unclear. To address this problem, we propose a multi-source prediction-powered inference method by aggregating multiple pseudo-labeled datasets together, where the aggregation weights are estimated by minimizing the asymptotic volume of the resulting confidence region. We study both homogeneous settings, where the source and target distributions coincide, and heterogeneous settings, where distributional discrepancies arise between source and target distributions, including covariate shift and domain shift. Theoretically, we establish the asymptotic normality of the proposed estimator and show that the resulting confidence-region volume is asymptotically equivalent to the oracle optimal volume within the proposed weighting class. We further characterize when our method yields smaller confidence regions compared with both classical target-only inference and single-source prediction-powered inference. Simulation studies and a real-data application on dual-energy X-ray absorptiometry measured high body fat prevalence show that MPPI can reduce confidence-region volume while maintaining inferential validity in the settings considered.
\end{abstract}

\noindent%
\vfill

\newpage
\spacingset{1.8} 

\section{Introduction}

Statistical inference is challenged by the limited availability of high-quality labeled data and the growing complexity of modern datasets. 
In many applications, only a small gold-standard labeled dataset is available for the target task, which we denote as the target dataset. Meanwhile, much larger collections of unlabeled related datasets can often be accessed from previous studies or other repositories \citep{chapelle2006ssl}, which we refer to as the source datasets. When the target sample size is small, classical inference based solely on the target data often suffers from high variability, resulting in wide confidence regions. Leveraging information from related source datasets is therefore a natural way to improve inference. A straightforward solution is to pool the source and target datasets, but this is often infeasible because the source labels may be unavailable due to the cost of labeling or privacy constraints. One practical strategy is to leverage machine learning methods that can capture complex data structures and generate pseudo-labels without requiring heavy human annotation. For example, AlphaFold predicts atom-level three-dimensional protein structures from amino-acid sequences, providing pseudo-labels for proteins whose structures have not been experimentally determined \citep{Jumper2021AlphaFold}. Similarly, convolutional neural networks can predict nighttime light intensity from daytime satellite imagery, generating pseudo-labels for local economic activity in regions where direct measurements are unavailable \citep{Jean2016PovertySatellite}.
While machine-learning-generated pseudo-labels are becoming increasingly available across diverse domains \citep{tunyasuvunakool2021highly,lin2023evolutionary,zheng2023chatgpt}, developing statistically efficient inference procedures that can effectively exploit source information remains an important challenge.

Our goal is to leverage multiple pseudo-labeled source datasets together with a gold-standard labeled target dataset to improve inference on a target parameter. A simple but naive idea is to directly augment the target data with these pseudo-labeled datasets to increase the effective sample size. However, such pooling can be poor because the pseudo-labels may be inaccurate, source datasets may exhibit distributional mismatch with the target population, and different source datasets may vary substantially in their usefulness for inference on the target parameter. These challenges call for more principled approaches that can leverage pseudo-labeled data while maintaining valid statistical inference. An important recent line of work has focused on developing valid inference procedures based on machine-learning-generated pseudo-labels. A few representative methods incorporate pseudo-labeled data while preserving asymptotically valid confidence regions, including prediction-powered inference (PPI, \citet{Anastasios2023science}) and its extensions \citep{angelopoulos2024PPIplus,zrnic2024cppi,shan2026sadasafeadaptiveaggregation}. Nonetheless, existing PPI-style methods are primarily developed for the single-source setting and are mostly studied under homogeneous distributions, with relatively few results available for heterogeneous settings. Their theoretical analyses mainly focus on asymptotic validity. How to combine multiple pseudo-label score sources under potential distributional mismatch while maintaining validity and achieving tighter confidence regions remains largely unexplored. Therefore, our paper focuses on the following question.

\textit{How can we calibrate and aggregate multiple pseudo-label score sources so that the resulting confidence region remains asymptotically valid and as tight as possible?}

This paper addresses the above question via a multi-source prediction-powered inference (MPPI) framework, which chooses data-driven weights to directly minimize the volume of the resulting confidence region.  
Suppose that we have a labeled target dataset and $S$ unlabeled source datasets, each equipped with a pretrained machine learning method. For each source dataset $1\le s \le S$, we first construct a modified empirical risk that adjusts the pseudo-labels using the target sample. We then assign data-driven weights to the target empirical risk and the $S$ modified empirical risks, to construct an estimator of the target parameter. The weights are then selected to minimize the estimated volume of the resulting confidence region. With estimated weights, we obtain our MPPI estimator. In both homogeneous and heterogeneous regimes, we establish that the MPPI estimator is asymptotically normal and that the resulting confidence region has volume asymptotically equivalent to that of the oracle optimal weighting rule.

We begin with the homogeneous regime, where the source and target data share the same distribution. 
This baseline setting removes distribution shift and matches the standard setup in prior PPI analyses, allowing us to focus on when and why multiple sources of pseudo-labeled information can improve inference. We first compare MPPI with the classical estimator based only on the target data, where single-source PPI is included as a special case. We show that MPPI can yield tighter confidence regions when the averaged pseudo-label directional score is sufficiently aligned with the true-label directional score. Next, we compare MPPI directly with single-source PPI, showing that MPPI can further improve inference when the averaged pseudo-label directional score provides a more accurate approximation to the true-label directional score while the additional variability from combining multiple auxiliary sources remains well controlled.

We next extend MPPI to heterogeneous regimes, where the source and target distributions differ, including under covariate shift and domain shift. Under covariate shift, the conditional distribution of responses given predictors is invariant across domains, while the marginal distribution of predictors changes.
Under domain shift, the source and target distributions are linked through a measure-preserving transformation. Most of the related literature under these two shift models focuses on prediction and predictive uncertainty quantification. For example, weighted conformal prediction under covariate shift has been studied in \citet{Tibshirani2019NIPS,qin2025distribution,jin2025model}, and \citet{ge2024optimal} develops aggregation methods for prediction intervals under unsupervised domain shift. In contrast, we focus on inference for general parameters within the MPPI framework. In both regimes, MPPI requires estimating alignment nuisance objects, such as density ratios under covariate shift \citep{sugiyama2007covariate,gretton2009covariate} and transport transformations under domain shift \citep{seguy2017large,makkuva2020optimal,divol2022unbalanced,deb2021wasserstein,yuan2025optimal}.
To account for the additional nuisance estimation required for distribution alignment, we estimate these objects on held-out folds and incorporate them into the MPPI estimating equations via cross-fitting. This separates nuisance estimation from target-parameter inference and facilitates valid asymptotic inference.

Our main contributions can be summarized as follows. Methodologically, we provide a unified multi-source framework that combines a gold-standard target dataset with multiple pseudo-labeled datasets via optimal aggregation. The framework covers a homogeneous setting and two heterogeneous settings including covariate shift and domain shift, and it selects weights by directly minimizing the volume of the confidence region.
Theoretically, our contributions are threefold. First, in each setting, we establish the asymptotic normality of the MPPI estimator. Second, we show that the feasible confidence-region volume is asymptotically equivalent to the oracle optimal confidence-region volume within the proposed weighting class. Third, we provide theoretical comparisons between MPPI, PPI, and classic inference. Specifically, we give interpretable sufficient conditions for MPPI to improve over the classical estimator and single-source PPI. These conditions clarify the roles of score alignment, approximation error, and variance cost in determining when multiple pseudo-label score sources are beneficial.
Numerically, we demonstrate MPPI's efficiency in extensive simulations and a real-data application to inference on DXA-measured high body fat prevalence.

The rest of the paper is organized as follows. Section~\ref{sec:methodology} presents the general MPPI framework. Section~\ref{sec:homogeneous} develops the homogeneous theory, and Section~\ref{sec:heterogeneous} extends the methodology and theory to covariate shift and domain shift.   Section~\ref{sec:sim} reports simulation evidence and comparisons with benchmarks.
Section~\ref{sec:real_data} presents a real-data application on DXA-measured body fat percentage.
Finally, we discuss the scope of PPI-type methods and future work in Section \ref{sec:conc}.
Detailed algorithms, additional numerical results, technical lemmas, and proofs are in the supplementary material.

\section{Model Setup}\label{sec:methodology}

Let $(X^{(0)},Y^{(0)}) \sim \mcP$ denote a generic observation from the target distribution. We are interested in the $p$-dimensional parameter defined as the unique minimizer
\begin{align}\label{S1}
    \btheta^{*} = \argmin_{\btheta \in \Theta} \mE\big[\ell_{\btheta}(X^{(0)}, Y^{(0)})\big],
\end{align}
where $\ell_{\btheta}(\cdot,\cdot)$ is a convex loss function in $\btheta$, and $\Theta \subset \mathbb{R}^p$ is a convex parameter space.
Let $\nabla \ell_{\btheta}(x,y)$ denote a subgradient of $\ell_{\btheta}$ with respect to $\btheta$. Equivalently,  $\btheta^*$  is characterized by the estimating equation $\mE(\nabla \ell_{\btheta^{*}}(X^{(0)}, Y^{(0)})) = 0$. 

Suppose that the gold-standard target dataset $\mcD^{(0)}$ contains observations $(\xizero, y_i^{(0)})$ for $i=1,\ldots,N_0$, where $(\xizero, y_i^{(0)})$ are i.i.d.s. drawn from $\mcP$.  Define the empirical risk on target data by $\widehat{R}^{(0)}_{\btheta} = N_0^{-1} \sum_{i=1}^{N_0}  \ell_{\btheta}(\xizero, y_i^{(0)})$. The classical estimator based on the target sample is  $\widehat{\btheta}^{(0)} \in \argmin_{\btheta \in \Theta} \widehat{R}^{(0)}_{\btheta}$.
Under standard regularity conditions, $\widehat{\btheta}^{(0)}$ is asymptotically normal and yields a Wald-type confidence region. However, when $N_0$ is small, inference of the classical estimator may be statistically inefficient, resulting in large confidence regions.

We consider $S$ source datasets $\mcD^{(1)},\ldots,\mcD^{(S)}$, each independent of the target dataset and of the other source datasets. The $s$-th source dataset consists of $N_s$ i.i.d. observations $(\xis,\yis)$ for $i=1,\ldots,N_s$ drawn from the source population $\mcQ^{(s)}$, together with a pretrained machine learning method $f^{(s)}$, where $f^{(s)}(x)$ denotes the pseudo-label produced at covariate value $x$. In practice, source labels $\yis$ are often unavailable to the target analyst, either because the source data are unlabeled or because labels cannot be shared due to privacy constraints. Moreover, the source distribution $\mcQ^{(s)}$ may differ from the target population $\mcP$, which necessitates domain alignment before source information can be used for valid inference. Given $f^{(s)}$, we construct the $s$-th pseudo-labeled source dataset by pairing the source covariates $\xis$ with pseudo-labels $f^{(s)}(\xis)$.

Our goal is to improve statistical inference for the target parameter by leveraging multiple pseudo-labeled source datasets together with the gold-standard labeled target dataset. This is challenging for several reasons. First, pseudo-labeled source information can be biased, due to errors in the pseudo-labels generated by $\fs$. Second, the source and target distributions may differ, so that source information may not be directly transferable to the target inferential problem and can induce an additional source-target gap. Third, due to privacy or data-sharing constraints, the target analyst may not be able to directly access and pool the source datasets with the target data. To address these issues, in the following content, we first develop the basic methodology under a homogeneous setting where the source and target distributions coincide. We then develop the framework to heterogeneous settings, including covariate shift and domain shift.

\section{MPPI under Homogeneous Distributions}\label{sec:homogeneous}
Suppose that the target dataset and all source datasets share the same data distribution, i.e., $\mcQ^{(1)} =  \ldots  = \mcQ^{(S)} = \mcP$. We assume that the $S$ machine learning methods $\fs$ are pretrained and provide a detailed methodological description and theoretical analysis.

\subsection{Methodology}
For each source dataset $1 \le s \le S$, we induce a modified empirical risk
\begin{align*}
\widehat{\text{MR}}_{\btheta}^{(s)} =  \frac{1}{N_s}\sum_{i=1}^{N_s} \ell_{\btheta}\left(\xis,\fs(\xis)\right) + \frac{1}{N_0}\sum_{i=1}^{N_0}\Bigl\{
\ell_{\btheta}(\xizero,y_i^{(0)})-\ell_{\btheta}\left(\xizero,\fs(\xizero)\right)
\Bigr\},
\end{align*}
where the first term $N_s^{-1}\sum_{i=1}^{N_s} \ell_{\btheta} (\xis,\fs(\xis) )$ evaluates the loss using pseudo-labels $\fs(\xis)$ in the $s$-th dataset. Since pseudo-labels may be biased relative to the gold-standard labels, the second term calibrates the pseudo risk using the target sample by adding the empirical gap between the gold-standard loss and the pseudo-label loss on the target covariates. Consequently, under the homogeneous distribution, the modified empirical risk has the same population target risk as the gold-standard empirical risk, while retaining source-side information through the pseudo-label component.

Let $\mathcal W$ be the simplex of nonnegative $(S+1)$-dimensional weight vectors whose components sum to one. Given weights $\bw=(w_0,\ldots,w_S)^\top\in\mathcal W$, we define the weighted estimator by
\begin{align}
 \widehat{\btheta}(\bw)
 \in
 \operatorname*{arg\,min}_{\btheta \in \Theta}
 \left[
 w_0 \widehat{R}^{(0)}_{\btheta}
 +
 \sum_{s=1}^S w_s \widehat{\text{MR}}_{\btheta}^{(s)}
 \right].
 \label{eq:mw+Deltaw}
\end{align}
From \eqref{eq:mw+Deltaw}, each source contributes through a calibrated approximation to the target risk. The resulting objective function combines the empirical target risk with multiple source-derived risk approximations, allowing information from different machine learning methods to be integrated into a unified estimation procedure. The target risk $\widehat R_{\btheta}^{(0)}$ is retained as a benchmark component, ensuring that the estimator remains anchored to the gold-standard labeled target data. The weight vector $\bw$ determines how the target and source information are balanced, with larger weights assigning greater influence to the corresponding risk components in the estimation of $\btheta^*$.

To better understand the role of $\hbtheta(\bw)$, we discuss it from two different perspectives. We first show some special cases covered in our framework. We then compare MPPI with other weighting-based approaches, whose weighting object is different from ours. We begin with the connection to existing PPI estimators.  If $w_0=1$ and all other components are zero, then $\hbtheta(\bw)$ reduces to the classical target-only estimator $\hbtheta^{(0)}$. For any $1\le s\le S$, if $w_s=1$ and all other components are zero, then $\hbtheta(\bw)$ reduces to the single-source estimator $\hbtheta^{(s)}$, which corresponds to a PPI$++$ estimator with the tuning parameter fixed at $1$. Therefore, our formulation can be viewed as a unified weighted extension of these existing PPI estimators.

We next compare our estimator with other weighting-based approaches including parameter averaging and predictor averaging. Although these methods also involve weighted combinations, the object being weighted is different.  The first weighting strategy is parameter averaging. This approach computes the estimators $\hbtheta^{(0)},\hbtheta^{(1)},\ldots,\hbtheta^{(S)}$ separately and then forms the weighted combination
$\sum_{s=0}^S w_s \hbtheta^{(s)}$,  following the general principle of frequentist model averaging \citet{wan2010least,Zhang2016jasa}. Instead, our estimator $\hbtheta(\bw)$ averages the empirical criteria before optimization and then solves the resulting problem. Hence, parameter averaging weights the minimizers, whereas our method weights the objective functions. This distinction is important because our weighted objective remains a direct sample analogue of a population risk, so the estimator continues to admit a clear population-level interpretation. By contrast, the parameter averaging $\sum_{s=0}^S w_s \hbtheta^{(s)}$ is generally not itself obtained by minimizing a corresponding population risk. The two constructions therefore differ in general, although they coincide in some simple special cases such as mean estimation under squared-error loss.

The second benchmark is predictor averaging. To illustrate the idea, consider a single source dataset $\mcD^{(1)}$ together with $M$ specified predictors $f^{(1)},\ldots,f^{(M)}$. Predictor averaging first combines these predictors into a single aggregated predictor and then plugs the aggregated predictor into the modified empirical criterion.\footnote{Define
$\widehat{\text{MR}}_{\btheta}(f)
=
N_1^{-1}\sum_{i=1}^{N_1}
\ell_{\btheta}\bigl(x_i^{(1)},f(x_i^{(1)})\bigr)
+
N_0^{-1}\sum_{i=1}^{N_0}
\{
\ell_{\btheta}\bigl(\xizero,y_i^{(0)}\bigr)
-
\ell_{\btheta}\bigl(\xizero,f(\xizero)\bigr)
\}$.
Let $f_{\bw}(x)=\sum_{m=1}^M w_m f^{(m)}(x)$. Then predictor averaging solves $\argmin_{\btheta\in\Theta}\widehat{\text{MR}}_{\btheta}(f_{\bw})$,
whereas our method solves $\argmin_{\btheta\in\Theta}\sum_{m=1}^M w_m \widehat{\text{MR}}_{\btheta}(f^{(m)})$.} Our method uses the weights differently. Instead of averaging the predictors first, we first construct the modified empirical criterion associated with each predictor and then average these criteria. Thus, predictor averaging operates at the predictor level, whereas our method operates at the criterion level. The two procedures are generally not equivalent. They can yield the same argmin only in special cases, for example when the part of the loss that depends on $\btheta$ is affine in the label, as in mean estimation or linear regression under squared-error loss, or when all predictors produce identical outputs. In the special case of mean inference, \citet{shan2026sadasafeadaptiveaggregation} can also be viewed as using a predictor-averaging form in the homogeneous labeled--unlabeled setting.

Having clarified the role of $\bw$ and its relation to existing methods, we now turn to weight selection. Our weights are chosen to tighten the confidence region of the resulting estimator. To construct the confidence region, we need a sample estimator of the asymptotic covariance matrix of $\hbtheta(\bw)$. 
To formalize this idea, we first introduce the relevant population-level quantities and then define their sample counterparts.  
Denote the covariance matrix of $\nabla \ell_{{\btheta}}(X^{(0)}, Y^{(0)})$ by $\bSig_{\btheta}$, $\nabla \ell_{{\btheta}}(X^{(0)},\fs(X^{(0)}))$ by $\bSig^{(s)}_{m_{\btheta}}$, and the covariance matrix of the rectifier  by $\bSig_{\Delta_{\btheta}}$, whose $(s,k)$ block is $\bSig^{(s,k)}_{\Delta_{\btheta}}$. Here $\bSig^{(s,k)}_{\Delta_{\btheta}}$ is the covariance between $\nabla \ell_{\btheta}(X^{(0)}, Y^{(0)}) - \nabla \ell_{\btheta} (X^{(0)}, \fs(X^{(0)}) )$ and $\nabla \ell_{\btheta}(X^{(0)}, Y^{(0)}) - \nabla \ell_{\btheta} (X^{(0)},  {f}^{(k)}(X^{(0)}) )$ for $1\le s,k \le S$. Define $\bSig_{m_{\btheta},\Delta_{\btheta}} = \bSig_{\Delta_{\btheta}}  +  \diag(c_1\bSig^{(1)}_{m_{\btheta}},\ldots,c_S\bSig^{(S)}_{m_{\btheta}})$,  where $ N_0/N_s \rightarrow c_s$. Define $\bC_{\btheta}$ as the block column whose $s$-th block is $\bSig_{\Delta_{\btheta}}^{(s,0)}$, where $\bSig_{\Delta_{\btheta}}^{(s,0)}$ is the covariance matrix of $\nabla \ell_{\btheta}(X^{(0)}, Y^{(0)}) - \nabla \ell_{\btheta} (X^{(0)}, \fs(X^{(0)}) )$ and $\nabla \ell_{\btheta}(X^{(0)}, Y^{(0)})$. Define \(\bQ_{\btheta}\) as the $2\times 2$ block matrix with \(\bSig_{\btheta}\) and \(\bSig_{m_{\btheta},\Delta_{\btheta}}\) as the upper-left and lower-right blocks, respectively, and with \(\bC_{\btheta}^{\top}\) and \(\bC_{\btheta}\) as the upper-right and lower-left blocks, respectively.
Let  $\bA_{\btheta} =  \mathbb{E} (\nabla^2  \ell_{\btheta}(X^{(0)},Y^{(0)} ))$ where $\nabla^2$ denotes the Hessian matrix with respect to $\btheta$. Define $\bSig(\btheta,\bw) = \bA_{\btheta}^{-1} (\bw \otimes \bI_p)^{\top} \bQ_{\btheta}(\bw \otimes \bI_p) \bA_{\btheta}^{-1}$.
Throughout, the sample analogues of these quantities are denoted by placing a ``hat'' over the corresponding symbol, e.g., $\hbSig_{\btheta}$, $\widehat{\bC}_{\btheta}$, $\hbQ_{\btheta}$, and $\hbSig(\btheta,\bw) = \widehat{\bA}_{\btheta}^{-1}(\bw \otimes \bI_p)^{\top} \hbQ_{\btheta}(\bw \otimes \bI_p) \widehat{\bA}_{\btheta}^{-1}$.

We then construct the confidence region as
\begin{align}\label{eq:mppi_confidence_region}
C^{\mathrm{MPPI}}_{\alpha}(\bw) =\left\{
\btheta\in\Theta:
N_0\{\btheta-\widehat{\btheta}(\bw)\}^{\top}
\widehat{\bSig}^{-1}\{\widehat{\btheta}(\bw),\bw\}
\{\btheta-\widehat{\btheta}(\bw)\}
\le
\chi^2_{p,1-\alpha}
\right\},
\end{align}
where $\mathcal{B}_{\bSig,\alpha}$ denotes the ellipsoid defined by $\bh^{\top}\bSig^{-1} \bh \le \chi_{p, 1-\alpha}^2$, and $\chi^{2}_{p,1-\alpha}$ denotes the $(1-\alpha)$-quantile of the chi-square distribution with $p$ degrees of freedom. The construction procedure of $\mathcal{B}_{\widehat{\bSig}(\hbtheta(\bw), \bw)/N_0,\alpha}$ is presented in Section B of the Supplementary Material \citep{mppi2026} for page limit, and the corresponding Wald testing procedure is summarized in Section A of the Supplementary Material \citep{mppi2026}. The weights $\bw$ aim to minimize the volume of the confidence region, which refers to shortening the width of the confidence interval (CI) if the target estimand is one dimensional. A smaller confidence region means less uncertainty, and  implies more reliable estimates. Specifically, the volume of the confidence region is 
\begin{align}\label{eq:volume}
    \mathrm{Vol}(\mathcal{B}_{\widehat{\bSig}(\hbtheta(\bw), \bw)/N_0,\alpha} ) = \frac{\pi^{p/2} (\chi_{p,1-\alpha}^2)^{p/2} }{\Gamma(\frac{p}{2} + 1)} \frac{\sqrt{   \det\left(\widehat{\bSig}(\hbtheta(\bw), \bw)\right)  }}{N_0^{p/2}}.
\end{align}
Since only the term $\det (\hbSig(\hbtheta(\bw),\bw) ) $ in \eqref{eq:volume}  is related to $\bw$, to minimize \eqref{eq:volume} relative to $\bw \in \mathcal{W}$ equals to minimize $\det (\hbSig(\hbtheta(\bw),\bw) ) $ over $\bw \in \mathcal{W}$. 
Our weighting criterion is 
\begin{align}\label{eq:weight}
    \mcC_N (\bw) =  \log \det\left(\hbSig(\hbtheta(\bw),\bw)\right)  \text{ and } 
    \hbw \in \argmin_{\bw \in \mathcal{W}} \mcC_N (\bw).
\end{align} 
The above weighting criterion \eqref{eq:weight} is based on confidence regions. For completeness, we also provide a coordinate-wise Bonferroni construction for the MPPI estimator in Section B of the Supplementary Material \citep{mppi2026}.
Weighting criterion \eqref{eq:weight} aims to provide the tightest confidence region, and the model averaging estimator with estimated weights $\hbtheta(\hbw)$ is called MPPI estimator.  The target dataset plays two roles in \eqref{eq:weight}. First, it calibrates the source predictions via the rectifier, thereby correcting the discrepancy between source-based pseudo-labels and the gold-standard labels. Second, it enters the weighted objective directly through the weight $w_0$. In this way, the weighting scheme allows the target data to play both a corrective role and a direct inferential role, which differs from PPI$++$, where the target sample is used only for calibration through a tuning parameter.

To minimize the weighting criterion in \eqref{eq:weight}, we propose the general iterative procedure summarized in Algorithm~\ref{alg1}. For certain loss functions and target parameters, the covariance estimator \(\widehat{\bSig}(\widehat{\btheta}(\bw),\bw)\) can be simplified analytically, yielding more efficient problem-specific algorithms. Section D of the Supplementary Material \citep{mppi2026} presents two such examples and gives the corresponding computational details. In our numerical experiments, the algorithm converges rapidly, typically reaching the stopping criterion within only a few iterations. In addition, the optimization problem in \eqref{eq:weight} can be reformulated as a joint optimization problem that solves simultaneously for the weighted estimator and the weights. This alternative formulation can be handled efficiently by a proximal alternating minimization scheme \citep{bolte2014pam,attouch2010lpam}, which is detailed in Section D of the Supplementary Material \citep{mppi2026}.

In the iterative algorithm below, $\bv^t$ denotes the weight vector at iteration $t$, while $\hbw$ denotes the final estimated weight vector.
\begin{algorithm}[h] 
\caption{Weighting Under Homogeneous Distribution}
\label{alg1}
\KwIn{Target dataset $\mcD^{(0)}$ with observations $(\xizero, y_i^{(0)})$ for $i=1,\ldots,N_0$, $S$ source datasets $\mcD^{(s)}$ with observations $(\xis, \yis)$ for $i=1,\ldots,N_s$ and $1\le s \le S$, machine learning methods $\fs$, significance level $\alpha \in (0,1)$, tolerance $\varepsilon > 0$ (default $\varepsilon = 10^{-6}$), and maximum iterations $T_{\max}$ (default $T_{\max} = 1000$).}
\BlankLine
Initialize $\bv^0=\bigl((S+1)^{-1},\ldots,(S+1)^{-1}\bigr)^\top$. \\
Compute $\widehat{\btheta}(\bv^0)=\argmin_{\btheta\in\Theta}\bigl(v_0^0 \widehat{R}^{(0)}_{\btheta}+\sum_{s=1}^S v_s^0 \widehat{\text{MR}}_{\btheta}^{(s)}\bigr)$ and $\mcC_N(\bv^0)=\log\det\Bigl(\widehat{\bSig}\bigl(\widehat{\btheta}(\bv^0),\bv^0\bigr)\Bigr)$. \\
\For{$1 \le t \le T_{\max}$}{ 
    Update the weights by setting $\bv^{t} = \argmin_{\bv \in \mathcal{W}} \log\det(\hbSig(\hbtheta(\bv^{t-1}),\bv))$; \\ 
    Update the estimator $\widehat{\btheta}(\bv^t)=\argmin_{\btheta\in\Theta} \bigl(v_0^t \widehat{R}^{(0)}_{\btheta}+\sum_{s=1}^S v_s^t\widehat{\text{MR}}_{\btheta}^{(s)}\bigr)$;  \\
    Update the volume as $\mcC_N(\bv^t)= \log\det\Bigl(\widehat{\bSig}\bigl(\widehat{\btheta}(\bv^t),\bv^t\bigr)\Bigr)$;\\
  \If{$|\mcC_N(\bv^{t}) - \mcC_N(\bv^{t-1})| < \epsilon$}{$\hbw=\bv^{t}$; \textbf{break}}
}
 
\KwOut{MPPI weight vector $\hbw$.}
\end{algorithm}

\subsection{Theoretical Properties}

In this section, we assume that the target and source samples are i.i.d. draws from $\mcP$. We next impose a set of standard regularity conditions to support our asymptotic analysis.

\begin{assumption}\label{as:convex}
    The population risk $\mE(\ell_{\btheta}(X^{(0)},Y^{(0)}))$ admits a unique minimizer $\btheta^{*}$ over the compact parameter space $\Theta$, and  $\btheta^*$ is an interior point of $\Theta$. 
\end{assumption}
Assumption~\ref{as:convex} ensures that the target parameter is well defined and identifiable as the unique minimizer of the population risk. The interior-point condition places \(\btheta^*\) away from the boundary of $\Theta$, so that the local perturbations around \(\btheta^*\) can be analyzed using standard first-order and second-order  Taylor expansions.

\begin{assumption}\label{as:hessian}
There exists a neighborhood $\Theta^*$ of $\btheta^*$ such that $\Theta^*\subset\Theta$ and, for each $s=1,\ldots,S$. The loss functions $\ell_{\btheta}(X^{(0)},Y^{(0)})$ and $ \ell_{\btheta}(X^{(0)},\fs(X^{(0)}))$ are twice continuously differentiable with
respect to $\btheta \in \Theta^*$. $\mathbb{E}(\sup_{\btheta\in\Theta^*} \|\nabla \ell_{\btheta}(X^{(0)},Y^{(0)})\|^2) < \infty$, and $\mathbb{E}(\sup_{\btheta \in \Theta^*}\|\nabla \ell_{\btheta}(X^{(0)},\fs(X^{(0)})) \|^2) < \infty$.   $\mathbb{E}(\sup_{\btheta \in \Theta^*}\|\nabla^2 \ell_{\btheta}(X^{(0)},Y^{(0)}) \|^2) < \infty$, and $\mathbb{E}(\sup_{\btheta \in \Theta^*}\|\nabla^2 \ell_{\btheta}(X^{(0)},\fs(X^{(0)})) \|^2) < \infty$.
\end{assumption}
Assumption~\ref{as:hessian} is a local regularity condition on the score functions and Hessian matrices generated by both the true-label loss and the pseudo-label losses. Specifically, it requires the maps $\btheta\mapsto \ell_{\btheta}(X^{(0)},Y^{(0)})$ and $\btheta\mapsto \ell_{\btheta}(X^{(0)},f^{(s)}(X^{(0)}))$ to be twice continuously differentiable on a neighborhood $\Theta^*$ of $\btheta^*$, and requires the corresponding scores and Hessians to have uniformly bounded second moments over $\Theta^*$. These conditions allow us to apply a Taylor expansion to the estimating equation around $\btheta^*$ and to apply the multivariate central limit theorem to the score terms. For common smooth losses, these moment bounds hold when these score and hessian have finite second moments uniformly for $\btheta\in\Theta^*$.

\begin{assumption}\label{as:hessian_A}
The Hessian matrix $\bA_{\btheta}=\mathbb{E}(\nabla^2\ell_{\btheta}(X^{(0)},Y^{(0)}))$ is uniformly positive definite and uniformly bounded over $\btheta\in\Theta^*$.
\end{assumption}
Assumption~\ref{as:hessian_A} imposes a standard non-degeneracy condition on the $\bA_{\btheta}$. The positive lower bound on the eigenvalues ensures that the risk function is locally curved rather than flat around \(\btheta^*\), while the upper bound rules out excessively large curvature.

\begin{assumption}\label{as:speed}
    $N_0/N_s \rightarrow c_s$ with positive constants $c_s$ for $ 1 \le s \le S$.
\end{assumption}
Assumption~\ref{as:speed} specifies the limiting ratios between the target sample size and the source sample sizes. The condition $N_0/N_s\to c_s\in(0,\infty)$ means that each source sample size is of the same asymptotic order as the target sample size. Under the $\sqrt{N_0}$-normalization used for the final estimator, source empirical averages based on $N_s$ observations are therefore rescaled by the limiting factor $c_s$. This assumption provides the constants that appear in the limiting covariance expressions and excludes regimes where a source sample size is asymptotically negligible or infinitely large relative to $N_0$.

\begin{assumption}\label{as:cov_regular}
The population covariance matrix $\bSig(\btheta,\bw)$ is uniformly positive definite and uniformly bounded over $(\btheta,\bw)\in\Theta^*\times\mcW$.
\end{assumption}
Assumption~\ref{as:cov_regular} imposes a non-degeneracy condition on the covariance matrix. The uniform eigenvalue bounds require \(\bSig(\btheta,\bw)\) to remain positive definite and bounded for all \((\btheta,\bw)\in\Theta^*\times\mathcal W\). The uniform consistency of the plug-in covariance estimator is established in the supplementary proof under the same moment and uniform law conditions used for the estimating equations.

\begin{theorem}\label{theorem:MPPI-clt}
Suppose the target sample and the $S$ source samples are mutually independent, with each sample consisting of i.i.d. draws from $\mcP$. Under Assumptions \ref{as:convex}, \ref{as:hessian}, \ref{as:hessian_A}, and \ref{as:speed}, for any fixed $\bw \in \mcW$, we have $ \sqrt{N_0} (\hbtheta(\bw) - \btheta^{*}) \xrightarrow{d} \mcN (\bzero, \bSig(\btheta^{*},\bw))$. 
\end{theorem}
Theorem~\ref{theorem:MPPI-clt} establishes the asymptotic normality of the weighted estimator \(\widehat{\btheta}(\bw)\) for any fixed weight vector \(\bw\in\mathcal W\).  Theorem~\ref{theorem:MPPI-clt} includes several existing estimators as special cases. 
Let $\be_s$ denote the unit vector in $\mathbb R^{S+1}$ whose $s$-th component equals one and whose other components equal zero, with $s=0$ corresponding to the target sample. When $\bw=\be_0$, $\hbtheta(\bw)$ reduces to the classical estimator based only on the target sample. When $\bw=\be_s$ for some $1\le s\le S$, it reduces to the single-source PPI estimator based on the $s$-th machine learning method. More generally, when only $w_0$ and $w_s$ are nonzero, the objective takes a PPI$++$-type form with tuning parameter determined by the source weight $w_s$ under the simplex constraint.

Next, we consider to establish the asymptotic distribution of the MPPI estimator $\hbtheta(\hbw)$ with the estimated weights in \eqref{eq:weight}. Let us define $\bw^* \in \argmin_{\bw \in \mathcal{W}}  \log \det\left( \bSig(\btheta^*,\bw) \right) $ and suppose the minimizer is unique.

\begin{theorem}\label{theorem:CI_MA}
Suppose the assumptions of Theorem~\ref{theorem:MPPI-clt} and Assumption~\ref{as:cov_regular} hold and $\bw^*$ is unique, then the MPPI estimator with $\hbw$ in \eqref{eq:weight} satisfies
$\sqrt{N_0}\bigl(\hbtheta(\hbw)-\btheta^{*}\bigr)
\xrightarrow{d} \mcN\bigl(\bzero, \bSig(\btheta^{*},\bw^*)\bigr)$. 
\end{theorem}
Theorem~\ref{theorem:CI_MA} is the main inferential result for MPPI under homogeneous distributions. With the plug-in weight $\hbw$ selected by \eqref{eq:weight}, the MPPI estimator is asymptotically normal with covariance matrix $\bSig(\btheta^*,\bw^*)$, where $\bw^*$ is the oracle minimizer of the population log-determinant criterion.  This result provides the asymptotic normal approximation used to construct the MPPI confidence region and to analyze the plug-in determinant in the following corollaries.

\begin{corollary}\label{cor:MPPI_coverage}
Suppose the assumptions of Theorem~\ref{theorem:CI_MA} hold. Then the confidence region $C^{\mathrm{MPPI}}_{\alpha}(\hbw)$ in \eqref{eq:mppi_confidence_region} satisfies $\Pr(\btheta^*\in C^{\mathrm{MPPI}}_{\alpha}(\hbw))\to 1-\alpha$.
\end{corollary}
Corollary~\ref{cor:MPPI_coverage} establishes that the confidence region constructed from the MPPI estimator with plug-in weights $\hbw$ retains asymptotic coverage $1-\alpha$.
Beyond coverage, the plug-in covariance estimator also determines the volume criterion in \eqref{eq:volume} and the selected weight in \eqref{eq:weight}. The next result compares our implemented confidence region with the infeasible oracle confidence region.

\begin{corollary}\label{cor:plugin_oracle_det}
Suppose the assumptions of Theorem~\ref{theorem:CI_MA} hold. Then $\det(\hbSig(\hbtheta(\hbw),\hbw))=\det(\bSig(\btheta^*,\bw^*))(1+o_p(1))$.
\end{corollary}
Corollary~\ref{cor:plugin_oracle_det} compares the determinant of the plug-in covariance estimator with its oracle counterpart. Consequently, the plug-in confidence-region volume in \eqref{eq:volume} is asymptotically equivalent to the oracle volume associated with $\bw^*$. Hence, our data-driven MPPI method asymptotically attains the same confidence-region volume as the infeasible oracle weighting rule.

\subsection{Theoretical Comparisons among MPPI, PPI, and Classic Inference}
After establishing the asymptotic normality, coverage guarantee, and plug-in oracle determinant approximation for the MPPI estimator, we compare the proposed MPPI estimator with two benchmark procedures, the classical estimator that only uses the gold-standard target sample and the single-source PPI estimator that uses one source dataset with one machine learning method at a time. The comparison illustrates two mechanisms behind the efficiency gain of MPPI. First, by averaging multiple pseudo-label directional scores, MPPI can construct an aggregated pseudo-label directional score that is more aligned with the true-label directional score than any individual pseudo-label directional score. Second, by spreading weights across sources, MPPI can reduce the effective variance contribution of pseudo-labeled auxiliary data. These two effects jointly explain why the weighted estimator can achieve a smaller asymptotic covariance.

We introduce some notations first. For any $\bu \in \mathbb{R}^p$, define $G_{\btheta^*}(\bu)=\bu^{\top}\nabla \ell_{\btheta^*}(X^{(0)},Y^{(0)})$ and $G_{\btheta^*}^{(s)}(\bu)=\bu^{\top}\nabla \ell_{\btheta^*}(X^{(0)},f^{(s)}(X^{(0)}))$. Here, $G_{\btheta^*}(\bu)$ is the true-label directional score along $\bu$, whereas $G_{\btheta^*}^{(s)}(\bu)$ is the pseudo-label directional score obtained by replacing the target response with the pseudo-label generated by the $s$-th machine learning method $\fs$. For any fixed $\bw\in\mathcal W$, define $G_{\btheta^*}(\bu;\bw)=\sum_{s=1}^S w_s G_{\btheta^*}^{(s)}(\bu)$ as the weighted pseudo-label directional score. With these notations, we present the following theorems.

\begin{theorem}\label{cor:MPPI_classic_suff}
 Suppose that the assumptions of Theorem~\ref{theorem:CI_MA} hold.   For any fixed $\bw\in\mathcal W$,  define $c_{\btheta^*}(\bu;\bw)= {\sum_{s=1}^S c_s w_s^2 \var (G_{\btheta^*}^{(s)}(\bu) )}/
{\var (G_{\btheta^*}(\bu;\bw) )}$. Assume that for every nonzero $\bu\in\mathbb{R}^p$ satisfying $\var(G_{\btheta^*}(\bu;\bw))>0$, we have
\begin{align}\label{eq:rho-threshold-multi}
\rho(\bu;\bw)
=\frac{\cov \bigl(G_{\btheta^*}(\bu),G_{\btheta^*}(\bu;\bw)\bigr)}{\var \bigl(G_{\btheta^*}(\bu;\bw)\bigr)}
\ge \frac{1+c_{\btheta^*}(\bu;\bw)}{2}.
\end{align}
Then the MPPI estimator associated with the given weight vector $\bw$ has asymptotic covariance no larger than that of the classical estimator, namely
$\det(\bSig(\btheta^*,\bw)) \le \det(\bSig(\btheta^*,\be_0))$.
\end{theorem}

Theorem~\ref{cor:MPPI_classic_suff} shows that MPPI can improve over the classical estimator when the weighted pseudo-label directional score is sufficiently close to the true-label directional score. In \eqref{eq:rho-threshold-multi}, the projection coefficient $\rho(\bu;\bw)$ measures how strongly the weighted pseudo-label directional score $G_{\btheta^*}(\bu;\bw)$ aligns with the true-label directional score $G_{\btheta^*}(\bu)$ along direction $\bu$. The term $c_{\btheta^*}(\bu;\bw)$ reflects the price paid for using source samples. In particular, the source-side variance cost contains
$\sum_{s=1}^S c_s w_s^2\var(G_{\btheta^*}^{(s)}(\bu))$ because the empirical fluctuation from source $s$ is multiplied by $w_s$ before its variance is taken. When several useful sources have comparable values of
$c_s\var(G_{\btheta^*}^{(s)}(\bu))$, spreading a fixed amount of source weight across them reduces this numerator through the squared weight. Therefore, when a favorable weight vector moves $G_{\btheta^*}(\bu;\bw)$ closer to the direction of the true-label directional score while keeping the squared-weight variance cost small, MPPI improves over classic inference.

\begin{figure}[htbp]
  \centering
  \includegraphics[width=\linewidth]{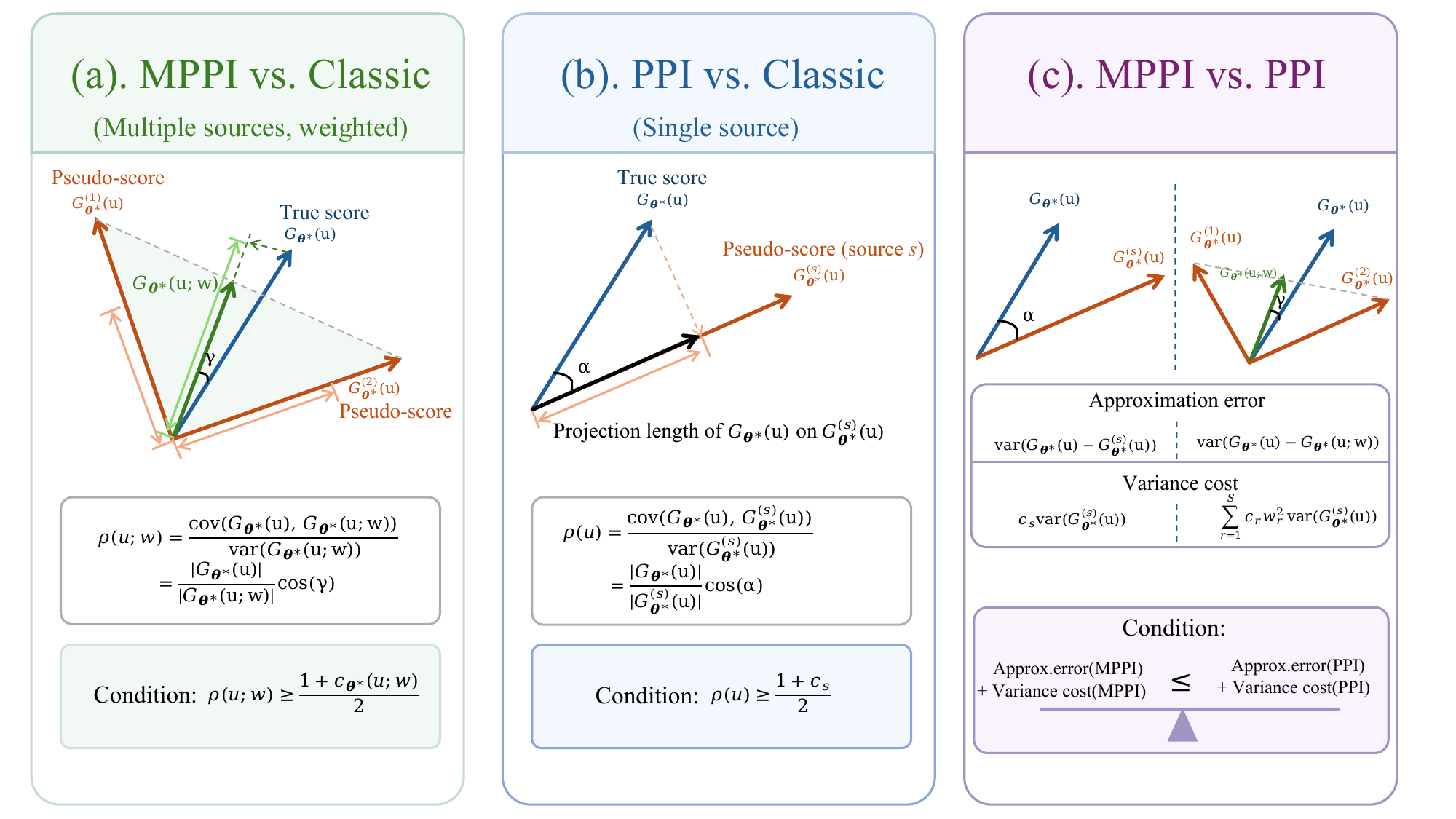}
  \caption{The comparisons among MPPI, PPI, and classic inference.}
  \label{fig:mppi_compare}
\end{figure}

Figure~\ref{fig:mppi_compare}(a) gives the corresponding geometric view of Theorem~\ref{cor:MPPI_classic_suff}. The weighted pseudo-label directional score $G_{\btheta^*}(\bu;\bw)$ lies in the scaled convex hull generated by the source pseudo-label directional scores $G_{\btheta^*}^{(s)}(\bu)$ for $1\le s\le S$, which is the green triangle space for $S=2$ in Figure~\ref{fig:mppi_compare}(b). For the left-hand side of \eqref{eq:rho-threshold-multi}, $\rho(\bu;\bw)$ can be written as
$\rho(\bu;\bw)=\|G_{\btheta^*}(\bu)\|\cos(\gamma)/\|G_{\btheta^*}(\bu;\bw)\|$,
where $\gamma$ is the angle between $G_{\btheta^*}(\bu)$ and $G_{\btheta^*}(\bu;\bw)$. 
Hence, by choosing $\bw$, MPPI can move the weighted pseudo-label directional score toward the direction of the true-label directional score even when individual pseudo-label directional scores are not perfectly aligned.

Theorem~\ref{cor:MPPI_classic_suff} covers the case of single-source PPI inference. To see this, take $\bw=\be_s$. Then $G_{\btheta^*}(\bu;\be_s)=G_{\btheta^*}^{(s)}(\bu)$ and $c_{\btheta^*}(\bu;\be_s)=c_s$. Hence, we have the following corollary.
\begin{corollary}  \label{cor:PPI_classic}
    Suppose the assumptions of Theorem~\ref{theorem:CI_MA} hold. Assume that, for every nonzero $\bu\in\mathbb R^p$, $\var(G_{\btheta^*}^{(s)}(\bu) )>0$ and 
\begin{equation}\label{eq:rho-threshold}
\rho(\bu) = \frac{\cov\!\left(G_{\btheta^*}(\bu),G_{\btheta^*}^{(s)}(\bu)\right)}
{\var\!\left(G_{\btheta^*}^{(s)}(\bu)\right)}
\ge \frac{1+c_s}{2}
\end{equation}
Then the single-source PPI estimator yields a tighter confidence region than the classical estimator, namely 
$ \det  (\bSig(\btheta^*,\be_s) )  \le   \det (\bSig(\btheta^*,\be_0) )$.
\end{corollary}

Corollary~\ref{cor:PPI_classic} states that the single-source PPI estimator improves over the classical estimator when the $s$-th pseudo-label directional score is sufficiently informative for the true-label directional score. This is a special case of Theorem~\ref{cor:MPPI_classic_suff} obtained by taking $\bw=\be_s$. In particular, Figure~\ref{fig:mppi_compare}(b) shows that $\rho(\bu)$ in \eqref{eq:rho-threshold} is the projection coefficient of the true-label directional score $G_{\btheta^*}(\bu)$ onto the pseudo-label directional score $G_{\btheta^*}^{(s)}(\bu)$. Equivalently,
$\rho(\bu)=\|G_{\btheta^*}(\bu)\|\cos(\alpha)/\|G_{\btheta^*}^{(s)}(\bu)\|$,
where $\alpha$ is the angle between these two directional scores. Thus, $\rho(\bu)$ quantifies how well the $s$-th pseudo-label directional score captures the true-label directional score along the direction $\bu$.
The right-hand side of \eqref{eq:rho-threshold}, $(1+c_s)/2$, is the required threshold for this single source. A larger source sample size relative to the target sample size leads to a smaller $c_s$ and hence a lower threshold, making the condition easier to satisfy. 

Comparing \eqref{eq:rho-threshold} with \eqref{eq:rho-threshold-multi} reveals the potential efficiency gain from aggregation in MPPI. As shown in Figure~\ref{fig:mppi_compare}(a), MPPI replaces the individual pseudo-label directional score $G_{\btheta^*}^{(s)}(\bu)$ by the weighted pseudo-label directional score $G_{\btheta^*}(\bu;\bw)$, which can lie closer to the direction of the true-label directional score when different pseudo-label directional scores provide complementary information. Therefore, even when a particular single-source PPI estimator has a small projection coefficient, or when several single-source PPI estimators fail to satisfy \eqref{eq:rho-threshold}, MPPI may still satisfy \eqref{eq:rho-threshold-multi} for a suitable weight vector, provided that the larger projection coefficient achieved by $G_{\btheta^*}(\bu;\bw)$ is not offset by the weighted source-sample variance cost.

Next, we directly compare  our weighted estimator and PPI methods to show when  our weighted estimator performs better than PPI.

\begin{corollary}\label{cor:MPPI_vs_PPI_direct}
Suppose the assumptions of Theorem~\ref{theorem:CI_MA} hold. For any $\bu\in\mathbb R^p$, define
$G_{\btheta^*}(\bu)=\bu^\top \nabla \ell_{\btheta^*}(X^{(0)},Y^{(0)})$ and  $G_{\btheta^*}^{(r)}(\bu) = \bu^\top \nabla \ell_{\btheta^*}(X^{(0)},f^{(r)}(X^{(0)}))$ for $r=1,\ldots,S$. For any fixed $\bw\in\mcW$, define $G_{\btheta^*}(\bu;\bw) = \sum_{r=1}^S w_r G_{\btheta^*}^{(r)}(\bu)$. Fix $s=1,\ldots,S$.  Assume that for every nonzero $\bu\in\mathbb R^p$,
\begin{align*}
\var (G_{\btheta^*}(\bu)-G_{\btheta^*}(\bu;\bw))
+\sum_{r=1}^S c_r w_r^2 \var (G_{\btheta^*}^{(r)}(\bu) )\le
\var (G_{\btheta^*}(\bu)-G_{\btheta^*}^{(s)}(\bu) )
+c_s \var (G_{\btheta^*}^{(s)}(\bu) ).
\end{align*}
Then $\det (\bSig(\btheta^*,\bw) ) \le \det (\bSig(\btheta^*,\be_s))$.
\end{corollary}

Corollary~\ref{cor:MPPI_vs_PPI_direct} gives a direct comparison between MPPI and single-source PPI. As illustrated in Figure~\ref{fig:mppi_compare}(c), the comparison can be understood through two components. The first component is the approximation error, which measures how close the pseudo-label directional score is to the true-label directional score. MPPI can reduce the approximation error when the individual pseudo-label directional scores contain complementary information and their weighted combination is closer to the true-label directional score than any single pseudo-label directional score. The second component is the variance cost induced by the auxiliary pseudo-labeled data. Importantly, the contribution of source $r$ enters through the squared weight $w_r^2$. Overall, Corollary~\ref{cor:MPPI_vs_PPI_direct} states that MPPI improves over the PPI estimator when there is a favorable balance between the reduction in approximation error and the variance cost of aggregating multiple sources. When the weighted pseudo-label directional score is more closely aligned with the true-label directional score and the weighted variance cost remains controlled, MPPI achieves a smaller asymptotic covariance than the single-source PPI estimator.

To better illustrate the above comparison results and their underlying mechanisms, we consider two examples that isolate different sources of gain. Example~\ref{ex:example1} concerns several pseudo-label directional scores that approximate the same nonlinear score-relevant component with different strengths, while Example~\ref{ex:example2} concerns pseudo-label directional scores that capture different signal components.
\begin{example}\label{ex:example1}
Consider a setting where the conditional mean is nonlinear but contains a linear component of interest, that is $Y=\beta^*X+X^2+\varepsilon$, where $X\sim N(0,1)$, $\varepsilon\sim N(0,\sigma^2)$, and $X\perp \varepsilon$. We define the inferential target as the coefficient in the linear working model $\beta X$, with $\beta\in\mathbb R$, under the squared loss $\ell_{\beta}(X,Y)=(Y-\beta X)^2$. In this example, the population risk under this working loss is strictly convex in $\beta$ since $\mathbb E(X^2)>0$, and its first-order condition is solved uniquely at $\beta=\beta^*$ because $\mathbb E(X^3)=0$ and $\mathbb E(X\varepsilon)=0$.

Suppose the $s$-th machine learning method takes the form $f^{(s)}(X)=\beta^*X+a_sX^2$. Then the asymptotic variances of the classical estimator, the $s$-th single-source PPI estimator, and the MPPI estimator are $\bSig_{\beta^*}^{\rm classic}=A_{\beta^*}^{-2}\,4(15+\sigma^2)$, $\bSig_{\beta^*}^{\rm PPI}(s)=A_{\beta^*}^{-2}\,4(((1-a_s)^2+c_s a_s^2)\cdot 15+\sigma^2)$, and $\bSig_{\beta^*}^{\rm MPPI}(\bw)=A_{\beta^*}^{-2}\,4(((1-\alpha(\bw))^2+\sum_{s=1}^S c_s w_s^2a_s^2)\cdot 15+\sigma^2)$ with $\alpha(\bw)=\sum_{s=1}^S w_sa_s$, respectively. 
    Therefore, if $0<a_s<2/(1+c_s)$, then the $s$-th single-source PPI estimator improves upon the classical estimator. If $(1-\alpha(\bw))^2+\sum_{s=1}^S c_s w_s^2a_s^2<1$, then the MPPI estimator improves upon the classical estimator. Moreover, MPPI improves upon the $s$-th single-source PPI whenever $(1-\alpha(\bw))^2+\sum_{r=1}^S c_r w_r^2a_r^2<(1-a_s)^2+c_s a_s^2$.

\end{example}
In Example \ref{ex:example1}, the inferential target is low-dimensional and linear, but the conditional mean contains the nonlinear component $X^2$. All source predictors approximate this same nonlinear component, with the source-specific coefficient $a_s$ determining its strength in the induced pseudo-label directional score. Thus, $\alpha(\bw)$ is the effective strength of this nonlinear score component after weighting. The term $(1-\alpha(\bw))^2$ measures the remaining mismatch in this score direction, while $\sum_{s=1}^S c_s w_s^2a_s^2$ is the source-sample variance cost. This example illustrates a calibration mechanism because MPPI can improve inference by choosing weights that make the aggregated nonlinear score component closer to its target strength while keeping the squared-weight variance cost controlled.

\begin{example}\label{ex:example2}
Consider inference on the population mean $\mu^*=\mathbb E(Y)$ under the squared loss $\ell_\mu(Y)=(Y-\mu)^2$. Suppose $Y=\mu^*+U_1+U_2+\varepsilon$, where $U_1$, $U_2$, and $\varepsilon$ are mutually independent, satisfy $\mathbb E(U_1)=\mathbb E(U_2)=\mathbb E(\varepsilon)=0$, $\var(U_1)=\var(U_2)=\tau^2$, and $\var(\varepsilon)=\sigma^2$. Since $\mathbb E\nabla \ell_\mu(Y)=-2\mathbb E(Y-\mu)$ and $A_{\mu^*}>0$, the population risk is strictly convex and has the unique minimizer $\mu^*$.

Suppose $X$ contains the two signal components $U_1$ and $U_2$, and the first and second machine learning methods recover them by $f^{(1)}(X)=\mu^*+U_1$ and $f^{(2)}(X)=\mu^*+U_2$. Assume also that $N_0/N_1\to c$ and $N_0/N_2\to c$. Then the asymptotic variances of the classical estimator, the single-source PPI estimators, and the MPPI estimator are $\bSig_{\mu^*}^{\rm classic}=A_{\mu^*}^{-2}\,4(2\tau^2+\sigma^2)$, $\bSig_{\mu^*}^{\rm PPI}(1)=\bSig_{\mu^*}^{\rm PPI}(2)=A_{\mu^*}^{-2}\,4((1+c)\tau^2+\sigma^2)$, and $\bSig_{\mu^*}^{\rm MPPI}(\bw)=A_{\mu^*}^{-2}\,4(((1-w_1)^2+(1-w_2)^2+c(w_1^2+w_2^2))\tau^2+\sigma^2)$, respectively. Therefore, if $c<1$, then each single-source PPI estimator improves upon the classical estimator. If $(1-w_1)^2+(1-w_2)^2+c(w_1^2+w_2^2)<2$, then the MPPI estimator improves upon the classical estimator. Moreover, MPPI improves upon either single-source PPI estimator whenever $(1-w_1)^2+(1-w_2)^2+c(w_1^2+w_2^2)<1+c$. 
\end{example}
Example \ref{ex:example2} illustrates a different mechanism. Here the two predictors capture different signal components, $U_1$ and $U_2$. A single-source PPI estimator can use only one component and leaves the other component in the residual score variation. In contrast, MPPI can assign positive weights to both sources and reduce the approximation error for both components, while the source-sample variance cost enters through the squared weights. For example, with $w_1=w_2=1/2$, the signal-related variance coefficient is $(1+c)/2$, compared with $1+c$ for each single-source PPI estimator and $2$ for the classical estimator. Thus, when $1<c<3$, each single-source PPI estimator has larger asymptotic variance than the classical estimator, whereas the equal-weight MPPI estimator has smaller asymptotic variance than the classical estimator. This example highlights complementarity across sources, together with the variance reduction induced by squared weights.

\section{MPPI under Heterogeneous Distributions}\label{sec:heterogeneous}

In this section, we extend MPPI to heterogeneous settings in which the $S$ source datasets are drawn from distributions $\mcQ^{(1)},\ldots,\mcQ^{(S)}$ that differ from the target distribution $\mcP$. Throughout the heterogeneous analysis, we rely on cross-fitting to separate the data used for distribution alignment and to compute the measure of fit and the rectifier.

\subsection{MPPI under Covariate Shift}
In this subsection, we consider a covariate-shift regime in which each source distribution $\mcQ^{(s)}$ differs from the target distribution $\mcP$ through a change in the marginal distribution of covariates, while the conditional distribution of responses given covariates remains invariant across domains. Specifically, we impose the following assumption.

\begin{assumption}\label{as:covariate_shift}
For each $s=1,\ldots,S$, $\mcQ^{(s)}$ is a covariate shift of $\mcP$, that is $\mcQ^{(s)}_{Y\mid X}=\mcP_{Y\mid X}$ and the density ratio is  
$\mathrm{DR}^{(s)}(x)={d\mcP_X}/{d\mcQ^{(s)}_X}(x)$.
\end{assumption}

\subsubsection{Methodology}
The source pseudo-risk can be reweighted to match the target covariate distribution.  If the density ratio $\mathrm{DR}^{(s)}(\cdot)$ is known, then for each $s=1,\ldots,S$, we define the reweighted modified empirical risk by
\begin{align*}
\widehat{\text{MR}}_{\btheta}^{\mathrm{DR},(s)}
=\frac{1}{N_s}\sum_{i=1}^{N_s}\mathrm{DR}^{(s)}(\xis)
\ell_{\btheta}\bigl(\xis,f^{(s)}(\xis)\bigr)
+\frac{1}{N_0}\sum_{i=1}^{N_0}\left\{\ell_{\btheta}(\xizero,\yizero)-\ell_{\btheta}\bigl(\xizero,f^{(s)}(\xizero)\bigr)
\right\}.
\end{align*}
Given weights $\bw=(w_0,\ldots,w_S)^\top\in\mcW$, the corresponding weighted estimator is defined by
$\hbtheta^{\mathrm{DR}}(\bw)
\in \argmin_{\btheta\in\Theta}
[w_0 \widehat{R}_{\btheta}^{(0)}
+\sum_{s=1}^S w_s \widehat{\text{MR}}_{\btheta}^{\mathrm{DR},(s)}
]$,
where the only change is the term $\mathrm{DR}^{(s)}$ compared with \eqref{eq:mw+Deltaw}. Let $\bSig_{m_{\btheta}}^{\mathrm{DR},(s)}$, $\bSig_{m_{\btheta},\Delta_{\btheta}}^{\mathrm{DR}}$, $\bQ_{\btheta}^{\mathrm{DR}}$, and $\bSig^{\mathrm{DR}}(\btheta,\bw)
=\bA_{\btheta}^{-1}(\bw\otimes\bI_p)^\top\bQ_{\btheta}^{\mathrm{DR}}(\bw\otimes\bI_p)\bA_{\btheta}^{-1}$
denote the corresponding population covariance quantities defined using the reweighted measure of fit. Define $\bw^{*,\mathrm{DR}} \in \argmin_{\bw\in\mcW} \log\det\bigl(\bSig^{\mathrm{DR}}(\btheta^*,\bw)\bigr)$ and assume $\bw^{*,\mathrm{DR}}$ is unique.

In practice, the density ratio $\mathrm{DR}^{(s)}(\cdot)$ is usually unknown and must be estimated from data. A variety of density-ratio estimation methods can be used here, including kernel mean matching \citep{huang2006NIPS}, the Kullback--Leibler importance estimation procedure \citep{Sugiyama2008AISM}, discriminative learning \citep{Bickel2009JMLR}, and many other approaches. To reduce overfitting and maintain the separation between density-ratio estimation and risk construction, we use cross-fitting. Specifically, fix the number of folds by $K$, and let $\mcI_k^{(0)}$ and $\mcI_k^{(s)}$ denote the index sets of the $k$-th fold in the target sample and the $s$-th source sample, respectively. For each $s$ and $k$, we estimate $\mathrm{DR}^{(s)}(\cdot)$ using the estimation covariates $\mcD^{(0)}\setminus \mcI_k^{(0)}$ and $\mcD^{(s)}\setminus \mcI_k^{(s)}$, and denote the resulting estimator by $\widehat{\mathrm{DR}}^{(s),-[k]}$. We then define the cross-fitted reweighted modified empirical risk by
\begin{align*}
\widehat{\text{MR}}_{\btheta}^{\widehat{\mathrm{DR}},(s)}
&=\frac{1}{K}\sum_{k=1}^K\frac{1}{|\mcI_k^{(s)}|}\sum_{i\in \mcI_k^{(s)}}\widehat{\mathrm{DR}}^{(s),-[k]}(\xis)
\ell_{\btheta}\bigl(\xis,f^{(s)}(\xis)\bigr)\notag\\
&\quad+\frac{1}{N_0}\sum_{i=1}^{N_0}\left\{\ell_{\btheta}(\xizero,\yizero)
-\ell_{\btheta}\bigl(\xizero,f^{(s)}(\xizero)\bigr)
\right\},
\end{align*}
and define the weighted estimator by
\begin{align*}
\hbtheta^{\widehat{\mathrm{DR}}}(\bw)
\in \argmin_{\btheta\in\Theta}\left[w_0 \widehat{R}_{\btheta}^{(0)}
+\sum_{s=1}^S w_s \widehat{\text{MR}}_{\btheta}^{\widehat{\mathrm{DR}},(s)}
\right]. 
\end{align*}

Let $\hbSig_{m_{\btheta}}^{\widehat{\mathrm{DR}},(s)}$, $\hbSig_{m_{\btheta},\Delta_{\btheta}}^{\widehat{\mathrm{DR}}}$, $\hbQ_{\btheta}^{\widehat{\mathrm{DR}}}$, and $\hbSig^{\widehat{\mathrm{DR}}}(\btheta,\bw)$ denote the corresponding sample analogues constructed from the cross-fitted reweighted measure of fit. Our weighting criterion under covariate shift is
\begin{align}\label{eq:weight_het}
\mcC_N^{\widehat{\mathrm{DR}}}(\bw)
=\log\det\Bigl(\hbSig^{\widehat{\mathrm{DR}}}(\hbtheta^{\widehat{\mathrm{DR}}}(\bw),\bw)\Bigr)
\text{ and } 
\hbw^{\widehat{\mathrm{DR}}}\in \argmin_{\bw\in\mcW}\mcC_N^{\widehat{\mathrm{DR}}}(\bw).
\end{align}

\subsubsection{Theoretical Properties} 
We impose the following assumptions to establish our theoretical results under covariate shift.
\begin{assumption}\label{as:stable_het}
    The number of sample splitting folds $K$ is fixed. For each
    $s=1,\ldots,S$ and $k=1,\ldots,K$, we have
    \begin{align*}
    \sup_{\btheta \in \Theta}\mE \left\{ \var_X \left(
    [\hRsk(X^{(s)})-\mathrm{DR}^{(s)}(X^{(s)})]
    \nabla \ell_{\btheta}(X^{(s)},\fs(X^{(s)}))
    \right) \right\}\rightarrow 0 ,
    \end{align*}
    and, on the neighborhood $\Theta^*$,
    \begin{align*}
    \sqrt{N_s}\sup_{\btheta\in\Theta^*}
    \left\|
    \mE_X\left(
    [\hRsk(X^{(s)})-\mathrm{DR}^{(s)}(X^{(s)})]
    \nabla \ell_{\btheta}(X^{(s)},\fs(X^{(s)}))
    \mid \hRsk
    \right)
    \right\|=o_p(1).
    \end{align*}
\end{assumption}
Assumption \ref{as:stable_het} requires $\hRsk$ to be stable around the true density ratio $\mathrm{DR}^{(s)}$, rather than around its own population mean. The first condition controls the conditional fluctuation induced by estimating the density ratio, and the second condition controls the corresponding score-level bias at the root-$N_s$ scale.  If the scale were only $O_p(1)$ rather than $o_p(1)$, an additional nuisance-induced bias or influence term could appear in the limiting distribution. The order $o_p(1)$ is used to make the estimated reweighted score first-order equivalent to the oracle reweighted score. If this term were only $O_p(1)$, the resulting estimator could still be consistent, but an additional nuisance-induced bias or influence term could remain in the limiting distribution. A similar distinction between consistency and oracle-type asymptotic normality  is also discussed in a recent work on covariate-shift transfer learning \citep{kato2024double}.
Assumption \ref{as:stable_het} is satisfied when the density ratio is known, when it is estimated with sufficiently high accuracy from external covariate data, or when a calibrated density-ratio estimator makes the displayed score bias negligible.

\begin{assumption}\label{as:ratio}
The  density ratio  $\mathrm{DR}^{(s)}$ satisfies  $ \mE[\{\mathrm{DR}^{(s)}(X^{(s)})\}^2]  < \infty$.  
\end{assumption}
Assumption~\ref{as:ratio} requires the density ratio to be square-integrable, which rules out severe covariate shift with insufficient overlap and prevents explosive importance weights.

\begin{assumption}\label{as:cov_regular_dr}
The population covariance matrix $\bSig^{\mathrm{DR}}(\btheta,\bw)$ is uniformly positive definite and uniformly bounded over $(\btheta,\bw)\in\Theta^*\times\mcW$.
\end{assumption}
Assumption~\ref{as:cov_regular_dr} is the covariate-shift analogue of Assumption~\ref{as:cov_regular}. It keeps the log-determinant criterion well-defined and stable under plug-in covariance estimation.

\begin{theorem} \label{theorem:CI_w_het}
Let the target observations $(x_i^{(0)},y_i^{(0)})$ for $i=1,\ldots,N_0$ be i.i.d. draws from $\mcP$, and let the source observations $(\xis,\yis)$ for $i=1,\ldots,N_s$ be i.i.d. draws from $\mcQ^{(s)}$ for each $s=1,\ldots,S$, where each $\mcQ^{(s)}$ is a covariate-shifted version of $\mcP$. All target and source datasets are assumed to be mutually independent.
Suppose Assumptions~\ref{as:convex}-\ref{as:speed} and \ref{as:covariate_shift}-\ref{as:cov_regular_dr} hold and $\bw^{*,\mathrm{DR}}$ is unique. Then we have
$\sqrt{N_0} (\hbtheta^{\widehat{\mathrm{DR}}} (\hbw^{\widehat{\mathrm{DR}}}) - \btheta^*) \xrightarrow{d} \mcN(0, \bSig^{\mathrm{DR}}(\btheta^*,\bw^{*,\mathrm{DR}}))$.
\end{theorem}
Theorem~\ref{theorem:CI_w_het} shows that after density-ratio alignment and cross-fitting, the MPPI estimator still converges to a normal limit with covariance matrix $\bSig^{\mathrm{DR}}(\btheta^*,\bw^{*,\mathrm{DR}})$. 
This theorem provides the distributional basis for the subsequent confidence-region and weight-comparison results. It justifies using the plug-in covariance matrix $\hbSig^{\widehat{\mathrm{DR}}}$ to construct Wald-type confidence regions for $\btheta^*$. It also shows that MPPI can still combine multiple pseudo-labeled sources for valid inference under covariate shift, provided that the density-ratio alignment is accurate enough at the score level. We next turn this normal approximation into an explicit coverage guarantee.

\begin{corollary}\label{cor:MPPI_coverage_covariate_shift}
Suppose the assumptions of Theorem~\ref{theorem:CI_w_het} hold and $\bSig^{\mathrm{DR}}(\btheta^*,\bw^{*,\mathrm{DR}})$ is positive definite. Let $C_{\alpha}^{\mathrm{MPPI,DR}}(\hbw^{\widehat{\mathrm{DR}}})$ be defined as in \eqref{eq:mppi_confidence_region}, with $\hbtheta(\hbw)$ and $\hbSig(\hbtheta(\hbw),\hbw)$ replaced by $\hbtheta^{\widehat{\mathrm{DR}}}(\hbw^{\widehat{\mathrm{DR}}})$ and $\hbSig^{\widehat{\mathrm{DR}}}(\hbtheta^{\widehat{\mathrm{DR}}}(\hbw^{\widehat{\mathrm{DR}}}),\hbw^{\widehat{\mathrm{DR}}})$, respectively. Then $\Pr(\btheta^*\in C_{\alpha}^{\mathrm{MPPI,DR}}(\hbw^{\widehat{\mathrm{DR}}}))\to 1-\alpha$.
\end{corollary}
Corollary~\ref{cor:MPPI_coverage_covariate_shift} turns the normal approximation in Theorem~\ref{theorem:CI_w_het} into a valid confidence region. Although the source covariate distributions differ from the target distribution, the density-ratio aligned MPPI procedure can still be used for Wald-type inference after accounting for the reweighted covariance structure. The next result compares the determinant used by the implemented procedure of MPPI under covariate shift with the determinant of the infeasible oracle one.

\begin{corollary}\label{cor:plugin_oracle_det_covariate_shift}
Suppose the assumptions of Theorem~\ref{theorem:CI_w_het} hold and $\bSig^{\mathrm{DR}}(\btheta^*,\bw^{*,\mathrm{DR}})$ is positive definite. Then $\det(\hbSig^{\widehat{\mathrm{DR}}}(\hbtheta^{\widehat{\mathrm{DR}}}(\hbw^{\widehat{\mathrm{DR}}}),\hbw^{\widehat{\mathrm{DR}}}))=\det(\bSig^{\mathrm{DR}}(\btheta^*,\bw^{*,\mathrm{DR}}))(1+o_p(1))$.
\end{corollary}
Corollary~\ref{cor:plugin_oracle_det_covariate_shift} shows that the plug-in determinant computed from $\hbSig^{\widehat{\mathrm{DR}}}$, $\hbtheta^{\widehat{\mathrm{DR}}}(\hbw^{\widehat{\mathrm{DR}}})$, and the selected weight $\hbw^{\widehat{\mathrm{DR}}}$ is asymptotically equivalent to the oracle determinant $\det(\bSig^{\mathrm{DR}}(\btheta^*,\bw^{*,\mathrm{DR}}))$. This result links the feasible weight-selection criterion in \eqref{eq:weight_het} to its population target.

\subsection{MPPI under Domain Shift}\label{sec:domain_shift}
We consider a general domain shift that can be characterized through a measure-preserving transport between covariate distributions. Define the covariate space of the $s$-th source dataset and the target dataset by $ \mcX^{(s)} $ and $\mcX^{(0)}$, respectively. Let $(X^{(s)},Y^{(s)})$ and $(X^{(0)},Y^{(0)})$ be sampled from $ \mcQ^{(s)} $ and $\mcP$, respectively. The domain shift structure assumption is presented below.

\begin{assumption}\label{as:domain_shift}
For each $1 \le s \le S$, there exists a measurable transport map
$\Ts$ from $\mcX^{(s)}$ to $\mcX^{(0)}$, i.e., $\Ts(X^{(s)}) \overset{d}{=} X^{(0)}$ such that
$\mcQ^{(s)}_{Y\mid X}(\cdot \mid X) =  \mcP_{Y\mid X}(\cdot \mid \Ts(X^{(s)}))$, for $ X^{(s)} \in \mcX^{(s)}$ a.e.. 
In addition, there exists a measurable map $\Ss:\mcX^{(0)}\to \mcX^{(s)}$ such that $ \Ss(X^{(0)}) \overset{d}{=} X^{(s)}$ and $\mE\| \Ss (\Ts(X^{(s)})) - X^{(s)} \| =o(1)$.
\end{assumption}
Assumption~\ref{as:domain_shift} describes domain shift through the measurable transport mapping. 
Compared with \citet{ge2024optimal}, we additionally assume the existence of a measurable map
$\Ts$ that transports the source covariate distribution back to the target,
together with $\mE\|\Ss(\Ts(X^{(s)}))-X^{(s)}\|=o(1)$.
This condition further requires an approximate invertibility that ensures that transferring functions or estimators defined on the target covariate space can be mapped back to the source space.
The condition holds, for example, when $\Ts$ is  one-to-one with a measurable inverse, or when $\Ts$ and $\Ss$ are estimated and the composition error vanishes as the sample size increases.

\subsubsection{Methodology}

For $1\le s\le S$, the source pseudo-risk can be aligned to the target covariate space through the transport map. For each $1\le s\le S$, define the modified empirical risk under domain shift by
\begin{align*}
\widehat{\text{MR}}_{\btheta}^{T,(s)}
&=\frac{1}{N_s}\sum_{i=1}^{N_s}\ell_{\btheta}\Bigl(T^{(s)}(\xis),f^{(s)}\bigl(S^{(s)}(T^{(s)}(\xis))\bigr)\Bigr)\\
&\quad+\frac{1}{N_0}\sum_{i=1}^{N_0}\left\{\ell_{\btheta}(\xizero,y_i^{(0)})
-\ell_{\btheta}\Bigl(\xizero,f^{(s)}(S^{(s)}(\xizero))\Bigr)\right\}.
\end{align*}
Given weights $\bw=(w_0,\ldots,w_S)^\top\in\mcW$, define the weighted estimator by
$\hbtheta^{T}(\bw)
\in \argmin_{\btheta\in\Theta} [w_0\widehat R_{\btheta}^{(0)}
+\sum_{s=1}^S w_s \widehat{\text{MR}}_{\btheta}^{T,(s)}
]$.
Analogously to the previous sections, let $\bQ_{\btheta}^{T}$ and
$\bSig^{T}(\btheta,\bw)$ denote the corresponding population covariance
quantities under domain shift. Define
$\bw^{*,T}\in\argmin_{\bw\in\mcW}\log\det(\bSig^{T}(\btheta^*,\bw))$
and assume $\bw^{*,T}$ is unique.

If $\Ts$ and $\Ss$ are not given, they can be estimated from the unlabeled source covariates and target covariates using standard domain alignment or optimal transport procedures   \citep{seguy2017large,makkuva2020optimal,divol2022unbalanced,deb2021wasserstein}. In our implementation, we estimate $\Ts$ jointly with an approximate right-inverse $\Ss$ by fitting a reverse map so that $\fs$ can be evaluated on the target covariates via $\fs(\Ss(x))$. Since the transport estimation can be challenging in high dimensions, we may also adopt simpler approximations that align low-order moments, such as correlation alignment (CORAL) \citep{sun2017coral}, or adversarial feature alignment methods such as adversarial discriminative domain adaptation (ADDA) \citep{tzeng2017adda}. Specifically, to estimate the $\Ts$ and $\Ss$ for $1\le s \le S$, we independently partition the $s$-th source dataset $\mcD^{(s)}$ and target dataset $\mcD^{(0)}$ into $K$ folds of equal size. Let $\mcI^{(s)}_k$  denote the index set of the  $k$-th fold in the  $s$-th source dataset, and $\mcI^{(0)}_k$ denote the index set of the $k$-th fold in the target dataset, for  $k=1,\ldots,K$. We use $\mcD^{(0)}  \setminus \mcI^{(0)}_k$ and $\mcD^{(s)}\setminus \mcI^{(s)}_k$ to estimate the transport maps $(\Ts,\Ss)$ and obtain the estimators $\hTsk$ and $\hSsk$. 

With estimators $\hTsk$ and $\hSsk$, we obtain
\begin{align*}
\widehat{\text{MR}}_{\btheta}^{\widehat T,(s)}
&=\frac{1}{K}\sum_{k=1}^K\frac{1}{|\mcI_k^{(s)}|}\sum_{i\in\mcI_k^{(s)}}\ell_{\btheta}\Bigl(\widehat T^{(s),-[k]}(\xis),f^{(s)}\bigl(\widehat S^{(s),-[k]}(\widehat T^{(s),-[k]}(\xis))\bigr)\Bigr)\notag\\
&\quad+\frac{1}{K}\sum_{k=1}^K\frac{1}{|\mcI_k^{(0)}|}\sum_{i\in\mcI_k^{(0)}}\left[\ell_{\btheta}(\xizero,y_i^{(0)})
-\ell_{\btheta}\Bigl(\xizero,f^{(s)}\bigl(\widehat S^{(s),-[k]}(\xizero)\bigr)\Bigr)
\right].
\end{align*}
Accordingly, define the weighted estimator
\begin{align*}
\hbtheta^{\widehat T}(\bw)\in \argmin_{\btheta\in\Theta}\left[w_0\widehat R_{\btheta}^{(0)}+\sum_{s=1}^S w_s \widehat{\text{MR}}_{\btheta}^{\widehat T,(s)}\right]. 
\end{align*}
Let $\hbQ_{\btheta}^{\widehat T}$ and $\hbSig^{\widehat T}(\btheta,\bw)$ denote the corresponding sample covariance quantities. The weights are chosen by
\begin{align}\label{eq:weight_domain}
\hbw^{\widehat T}\in \argmin_{\bw\in\mcW}\mcC_N^{\widehat T}(\bw)
=\argmin_{\bw\in\mcW}\log\det\Bigl(\hbSig^{\widehat T}(\hbtheta^{\widehat T}(\bw),\bw)\Bigr),
\end{align}
where $\hbSig^{\widehat T}(\btheta,\bw)=\widehat{\bA}_{\btheta}^{-1}(\bw\otimes\bI_p)^\top\hbQ_{\btheta}^{\widehat T}
(\bw\otimes\bI_p)\widehat{\bA}_{\btheta}^{-1}$. 

\subsubsection{Theoretical Properties}
We need the following assumptions to establish our theoretical results.

\begin{assumption}\label{as:stable_T}
The number of folds $K$ is fixed. For each $s=1,\ldots,S$ and
$k=1,\ldots,K$, assume that
\begin{align*}
&\sup_{\btheta\in\Theta}\mE\Big[\var_X\Big(
\nabla \ell_{\btheta}\big(\hTsk(X^{(s)}),\fs(\hSsk(\hTsk(X^{(s)})))\big) \\
&\qquad - \nabla \ell_{\btheta}\big(\Ts(X^{(s)}),\fs(\Ss(\Ts(X^{(s)})))\big) \Big)\Big] =o(1), \text{ and}\\
&\sup_{\btheta\in\Theta}\mE\Big[\var_X\Big(
\nabla \ell_{\btheta}\big(X^{(0)},\fs(\hSsk(X^{(0)}))\big)
- \nabla \ell_{\btheta}\big(X^{(0)},\fs(\Ss(X^{(0)}))\big)  \Big)\Big] =o(1).
\end{align*}
Moreover, on the neighborhood $\Theta^*$ of $\btheta^*$,
\begin{align*}
&\sqrt{N_s}\sup_{\btheta\in\Theta^*}
\Bigg\|
\mE_X\Big[
\nabla \ell_{\btheta}\big(\hTsk(X^{(s)}),\fs(\hSsk(\hTsk(X^{(s)})))\big)
\\
&\qquad\qquad
- \nabla \ell_{\btheta}\big(\Ts(X^{(s)}),\fs(\Ss(\Ts(X^{(s)})))\big)
\mid \hTsk,\hSsk
\Big]
\Bigg\|=o_p(1),\\
&\sqrt{N_0}\sup_{\btheta\in\Theta^*}
\Bigg\|
\mE_X\Big[
\nabla \ell_{\btheta}\big(X^{(0)},\fs(\hSsk(X^{(0)}))\big)
\\
&\qquad\qquad
- \nabla \ell_{\btheta}\big(X^{(0)},\fs(\Ss(X^{(0)}))\big)
\mid \hTsk,\hSsk
\Big]
\Bigg\|=o_p(1).
\end{align*} 
\end{assumption}
Assumption~\ref{as:stable_T} is a stability condition for the cross-fitted transport maps. It ensures that plugging in the cross-fitted transport map estimates $(\hTsk,\hSsk)$ in place of $(\Ts, \Ss)$ has a negligible impact on the gradient $\nabla \ell_{\btheta}(\cdot,\cdot)$, both through conditional fluctuations and through score-level bias at the root-sample-size scale. Compared with Assumption~\ref{as:stable_het}, Assumption~\ref{as:stable_T} has two conditional-variance conditions and two score-bias conditions because transport alignment affects two empirical components. The first component is the source pseudo-risk, where both $\hTsk$ and $\hSsk\circ\hTsk$ enter. The second component is the target-sample correction term, where the pseudo-label is evaluated through $\hSsk(X^{(0)})$. Thus, stability is required for both parts of the domain-shift estimating equation.

\begin{assumption}\label{as:hessian_domain}
There exists a neighborhood $\Theta^*\subset\Theta$ of $\btheta^*$. For each
$s=1,\ldots,S$, let $\mathcal{A}_s$ denote the collection of the three
input--label pairs
$(X^{(0)},Y^{(0)})$,
$(T^{(s)}(X^{(s)}), f^{(s)}(S^{(s)}(T^{(s)}(X^{(s)}))))$, and
$(X^{(0)}, f^{(s)}(S^{(s)}(X^{(0)})))$.
Then, for each $s=1,\ldots,S$ and every $(U,V)\in\mathcal{A}_s$, the loss $\ell_{\btheta}(U,V)$ is twice continuously differentiable with respect to $\btheta\in\Theta^*$, $\mathbb E(\sup_{\btheta\in\Theta^*}\|\nabla \ell_{\btheta}(U,V)\|^2)<\infty$ , $\mathbb E(\sup_{\btheta\in\Theta^*}\|\nabla^2 \ell_{\btheta}(U,V)\|^2)<\infty$. 
\end{assumption}
Assumption~\ref{as:hessian_domain} is the domain-shift analogue of Assumption~\ref{as:hessian}. The only difference is that the smoothness and moment requirements must now hold for all input--label pairs generated by the transport construction, including the target true-label pair, the transported source pseudo-label pair, and the target-side correction pair. These conditions ensure that the score CLT and the Taylor expansion around $\btheta^*$ remain valid under domain shift.

\begin{assumption}\label{as:cov_regular_t}
The population covariance matrix $\bSig^{T}(\btheta,\bw)$ is uniformly positive definite and uniformly bounded over $(\btheta,\bw)\in\Theta^*\times\mcW$.
\end{assumption}
Assumption~\ref{as:cov_regular_t} is the domain-shift analogue of Assumption~\ref{as:cov_regular}. It keeps the transport-based log-determinant criterion well-defined and stable under plug-in covariance estimation.

\begin{theorem} \label{theorem:CI_w_domain}
Let the target observations $(\xizero,\yizero)$ for $i=1,\ldots,N_0$ be i.i.d. draws from $\mcP$, and let the source observations $(\xis,\yis)$ for $i=1,\ldots,N_s$ be i.i.d. draws from $\mcQ^{(s)}$ for each $s=1,\ldots,S$, where each $\mcQ^{(s)}$ is a domain-shifted version of $\mcP$. All target and source datasets are assumed to be mutually independent.
Suppose Assumptions~\ref{as:convex}, \ref{as:hessian_A}, \ref{as:speed}, \ref{as:domain_shift}--\ref{as:hessian_domain}, and \ref{as:cov_regular_t} hold and $\bw^{*,T}$ is unique. Then we have
$\sqrt{N_0} (\hbtheta^{\widehat{T}} (\hbw^{\widehat{T}}) - \btheta^*) \xrightarrow{d} \mcN(0, \bSig^{T}(\btheta^*,\bw^{*,T}))$.
\end{theorem}

Theorem~\ref{theorem:CI_w_domain} establishes the asymptotic normality of the MPPI estimator with data-driven weights under domain shift. The result shows that, after aligning source covariates to the target covariate space through the transport maps and using cross-fitting to control the transport-estimation error, the MPPI estimator  still converges to a normal distribution with the limiting covariance $\bSig^{T}(\btheta^*,\bw^{*,T})$. 
This theorem gives the normal approximation needed for inference under domain shift. The next corollary turns this approximation into a coverage guarantee for the feasible transport-based confidence region.

\begin{corollary}\label{cor:MPPI_coverage_domain_shift}
Suppose the assumptions of Theorem~\ref{theorem:CI_w_domain} hold and $\bSig^{T}(\btheta^*,\bw^{*,T})$ is positive definite. Let $C_{\alpha}^{\mathrm{MPPI,T}}(\hbw^{\widehat T})$ be defined as in \eqref{eq:mppi_confidence_region}, with $\hbtheta(\hbw)$ and $\hbSig(\hbtheta(\hbw),\hbw)$ replaced by $\hbtheta^{\widehat T}(\hbw^{\widehat T})$ and $\hbSig^{\widehat T}(\hbtheta^{\widehat T}(\hbw^{\widehat T}),\hbw^{\widehat T})$, respectively. Then $\Pr(\btheta^*\in C_{\alpha}^{\mathrm{MPPI,T}}(\hbw^{\widehat T}))\to 1-\alpha$.
\end{corollary}

Corollary~\ref{cor:MPPI_coverage_domain_shift} shows that the confidence region constructed from $\hbtheta^{\widehat T}(\hbw^{\widehat T})$ and the plug-in covariance matrix $\hbSig^{\widehat T}$ has asymptotic coverage $1-\alpha$. This result is important because both the source pseudo-risk and the target correction term involve estimated transport maps. The corollary confirms that, under the stability conditions above, these additional estimation steps do not invalidate Wald-type inference for $\btheta^*$. 

\begin{corollary}\label{cor:plugin_oracle_det_domain_shift}
Suppose the assumptions of Theorem~\ref{theorem:CI_w_domain} hold and $\bSig^{T}(\btheta^*,\bw^{*,T})$ is positive definite. Then $\det(\hbSig^{\widehat T}(\hbtheta^{\widehat T}(\hbw^{\widehat T}),\hbw^{\widehat T}))=\det(\bSig^{T}(\btheta^*,\bw^{*,T}))(1+o_p(1))$.
\end{corollary}

Corollary~\ref{cor:plugin_oracle_det_domain_shift} shows that the determinant of the feasible plug-in covariance matrix is asymptotically equivalent to $\det(\bSig^{T}(\btheta^*,\bw^{*,T}))$.   

Beyond the transport-based type of domain shift in this subsection, one may allow a more general domain shift characterized through a common representation. Instead of requiring explicit transport maps between the source and target covariate spaces, suppose there exist measurable feature maps $\phi_s:\mcX^{(s)}\to \mcZ$ and $\phi_0:\mcX^{(0)}\to\mcZ$ such that $\phi_s(X^{(s)})\overset{d}{=}\phi_0(X^{(0)})$ and the conditional distribution is invariant given the representation, that is $\mcQ^{(s)}_{Y\mid \phi_s(X)}(\cdot\mid Z)=\mcP_{Y\mid \phi_0(X)}(\cdot\mid Z)$ for $Z\in\mcZ$. In this setting, MPPI can be constructed in the representation space by applying the estimating procedure to the transformed covariates. Under analogous smoothness, moment, and stability conditions in the representation space, the same asymptotic-normality argument can be adapted.
 
\section{Simulation Study}\label{sec:sim}
In this section, we conduct simulation studies to evaluate the empirical performance of MPPI. We consider a range of data-generating processes (DGPs) to assess both confidence region accuracy and hypothesis testing performance. We examine multiple inferential targets, each paired with its corresponding loss function. In the main text, we present the DGPs for mean inference. The corresponding linear regression parameter inference settings are described in Section E of the Supplementary Material \citep{mppi2026}. In this section, we use boldface notation to explicitly denote vectors and matrices for clarity in simulation and computation. In particular, $\bx_i$ represents a covariate vector and $\bX$ denotes the data matrix. 

\subsection{Data-Generating Process}

We consider one target dataset and three source datasets, that is $S=3$. Let $N_0$ and $N_s$ be the sample sizes of the target and $s$th source datasets, respectively. For $s=0,\dots,S$, let $\by^{(s)} = (y_1^{(s)},\dots,y_{N_s}^{(s)})^\top$ be the response vector and $\bX^{(s)} \in \mathbb{R}^{N_s \times d}$ be the design matrix with rows $\bx_i^{(s)} = (x_{i1}^{(s)},\dots,x_{id}^{(s)})^\top$ for $i=1,\dots,N_s$. We set the sample sizes to $N_0 = 5000$ for the target and $N_s = 2 N_0 s$ for the $s$th source ($s = 1, \dots, S$).

For homogeneous and covariate shift cases, suppose $\bx_i^{(0)} \overset{\text{i.i.d.}}{\sim} \mcN(\bm{0}_d,\bm{I}_d)$, and the true DGP is given by $y_i^{(0)} = g(\bx_i^{(0)}, \bbeta) + e_i^{(0)}$ with $e_i^{(0)} \overset{\text{i.i.d.}}{\sim} N(0,1)$ for $i = 1, \dots, N_0$. Set $d=2$ and $\bbeta = (1.2,-0.8)^{\top}$. We consider two DGPs, a linear model $g(\bx_i,\bbeta) = \bx_i^{\top}\bbeta$ and a nonlinear model $g(\bx_i,\bbeta) = 2x_{i1} + \sin(2\pi x_{i1}) + \sum_{j=2}^d \beta_jx_{ij}$. The mean target parameter is $\btheta^* = \mE(Y^{(0)})$, which equals $0$ under both the linear and nonlinear target models.

In the homogeneous setting, the source covariates and responses are generated from the same distribution as the target data. Specifically, $\bx_i^{(s)} \overset{\text{i.i.d.}}{\sim} \mcN(\bm{0}_d, \bm{I}_d)$, $y_i^{(s)} = g(\bx_i^{(s)}, \bbeta) + e_i^{(s)}$, and  $e_i^{(s)} \overset{\text{i.i.d.}}{\sim} N(0,1)$ for $i=1,\dots,N_s$ and $s=1,\dots,S$. Under covariate shift, the source and target domains share the same conditional response law given covariates, but may differ in their marginal covariate distributions. Specifically, $\bx_i^{(s)} \overset{\text{i.i.d.}}{\sim} \mcN(\bmu^{(s)}, \bm{I}_d)$, $y_i^{(s)} = g(\bx_i^{(s)}, \bbeta) + e_i^{(s)}$, and  $e_i^{(s)} \overset{\text{i.i.d.}}{\sim} N(0,1)$ for $i = 1, \dots, N_s$, $s=1,\dots,S$, where $\bmu^{(1)} =-0.5 \cdot \bm{1}_d$, $\bmu^{(2)} =-1 \cdot \bm{1}_d$, $\bmu^{(3)} =1.5 \cdot \bm{1}_d$, and $\bm{1}_d$ denotes the $d$-dimensional vector of ones. Since the density ratio is unknown, we estimate it using the Exponential Tilt–based density ratio method, as detailed in Section D.3 of the Supplementary Material \citep{mppi2026}.

We next consider the domain-shift setting. The target covariates are generated as $\bx_i^{(0)}\overset{\mathrm{i.i.d.}}{\sim}\mathcal{N}(\bzero_d,\bI_d)$. For each source $s=1,\ldots,S$, we first generate latent target-scale covariates
$\bz_i^{(s)}\overset{\mathrm{i.i.d.}}{\sim}\mathcal{N}(\bzero_d,\bI_d)$ and then construct the observed source-scale covariates by $\bx_i^{(s)}=\bA^{(s)}\bz_i^{(s)}+\bb^{(s)}$. Here $\bb^{(s)}=s \cdot(0.05,-0.025)^\top$, and $\bA^{(s)}=\operatorname{diag}(1+\alpha_s\bm q)$ with $\alpha_s=0.05 \cdot s$ and $\bm q=(0.5,1.0)^\top$. The target response is generated as $y_i^{(0)}=g_{\mathrm{M}}(\bx_i^{(0)})+e_i^{(0)}$, and the source responses are generated on the latent target scale as $y_i^{(s)}=g_{\mathrm{M}}(\bz_i^{(s)})+e_i^{(s)}$. We consider two DGPs. The linear DGP is $g_{\mathrm{M}}(\bz)=1+1.2z_1+1.2z_2$, and the nonlinear DGP is $g_{\mathrm{M}}(\bz)=1+1.2\sin(z_1)+1.2\sin(z_2)+0.4\sin(z_1+z_2)$. In this domain-shift mean setting, we use $e_i^{(s)}\sim N(0,7^2)$ for $s=0,\ldots,S$, and the target mean is $\btheta^*=\mE(Y^{(0)})=1$ under both DGPs.
The true source-to-target and target-to-source maps are $T_s(x)=(\bA^{(s)})^{-1}(x-\bb^{(s)})$ and $S_s(z)=\bA^{(s)}z+\bb^{(s)}$, respectively. Details of the affine transport estimators $\widehat{T}_s$ and $\widehat{S}_s$ are provided in Section D.4 of the Supplementary Material \citep{mppi2026}.

For pretrained machine learning models,  we generate an independent auxiliary sample with observations $(\bar{\bx}_i^{(s)},\bar{y}_i^{(s)})$ for $i=1,\ldots,N_s$ from each source dataset to train $f^{(s)}$, which is then fixed before applying all inferential procedures. In the homogeneous and covariate-shift settings, $(\bar{\bx}_i^{(s)},\bar{y}_i^{(s)})\overset{\mathrm{i.i.d.}}{\sim}\mathcal Q^{(s)}$. In the domain-shift mean setting, we generate
$\bar{\bz}_i^{(s)}\overset{\mathrm{i.i.d.}}{\sim}\mathcal N(\bzero_d,\bI_d)$, $\bar{\bx}_i^{(s)}=\bA^{(s)}\bar{\bz}_i^{(s)}+\bb^{(s)}$, and $\bar{y}_i^{(s)}=g_{\mathrm M}(\bar{\bz}_i^{(s)})+\bar e_i^{(s)}$ with $\bar e_i^{(s)}\overset{\mathrm{i.i.d.}}{\sim}N(0,7^2)$. Thus, $(\bar{\bx}_i^{(s)},\bar y_i^{(s)})$ are i.i.d. from the same joint distribution as $(\bx_i^{(s)},y_i^{(s)})$ for $i=1,\ldots,N_s$. To allow heterogeneous informativeness across sources, the machine learning methods $f^{(s)}$ are trained with different input features. Source~1 uses features derived from the first coordinate $\bar{x}_{i1}^{(s)}$, Source~2 uses features derived from the second coordinate $\bar{x}_{i2}^{(s)}$, and Source~3 uses only an intercept. For the nonlinear DGP, a sine transformation of the selected coordinate is additionally included for Sources~1 and~2.

\subsection{Competing Methods and Evaluation Criteria}
We compare MPPI with four baselines including   the classical estimator using only the target data,   PPI,  PPI++, and   an equal-weight (EW) variant of our method that assigns uniform weights to all datasets. Since PPI and PPI++ are defined for a single source dataset together with the target dataset, we report their performance separately for each of the three source datasets. For PPI++, we compute the tuning parameter $\widehat{\lambda}$ following Eq.~(8) of \citet{angelopoulos2024PPIplus}\footnote{Eq.~(8) of \citet{angelopoulos2024PPIplus} uses a preliminary estimator computed with any fixed $\lambda$ between $0$ and $1$. For source $s$, we use $\lambda=N_s/(N_0+N_s)$ as this fixed value.}. 
To clarify how MPPI differs from the classical estimator, PPI, and PPI++, Table~\ref{relationship} summarizes the corresponding weight allocations in each method, where our MPPI weights are data-driven.

\begin{table}[!ht]
    \centering
    \caption{Weight allocations in MPPI, classic estimator, PPI, and PPI++ when $S=3$}
    \begin{tabular}{lcccc}
    \toprule
    Method & Target & Source 1 & Source 2 & Source 3 \\
    \midrule
    MPPI & $\widehat w_0$ & $\widehat w_1$ & $\widehat w_2$ & $\widehat w_3$ \\
    Classic & $1$ & $0$ & $0$ & $0$ \\
    PPI (Source 1) & $0$ & $1$ & $0$ & $0$ \\
    PPI++ (Source 1) & $1-\widehat{\lambda}$ & $\widehat{\lambda}$ & $0$ & $0$ \\
    EW & $1/4$ & $1/4$ & $1/4$ & $1/4$ \\
    \bottomrule
    \end{tabular}
    \label{relationship}
\end{table}

To assess the reliability and efficiency of the constructed confidence regions, we focus on two key metrics including
the average coverage probability (ACP) and a generalized volume criterion (Vol). 
Specifically, over $B$ Monte Carlo replicates, we construct a $(1-\alpha)$ confidence region $\mcC_{\alpha}^b$ for $\btheta^*$ in the $b$-th replicate and compute $\text{ACP}=B^{-1}\sum_{b=1}^B \mathbb{I}(\btheta^* \in \mcC_{\alpha}^b)$ and $\text{Vol}=B^{-1}\sum_{b=1}^B \det (\widehat{\bSig}_b) $, where $\widehat{\bSig}_b$ denotes the estimated asymptotic covariance matrix in the $b$-th replicate. We set $B=1000$ to obtain stable estimates with reasonable computational cost. We also provide diagnostic evidence for the finite-sample normal approximation in Section F of the Supplementary Material \citep{mppi2026}.

In addition, we study the hypothesis testing problem for the mean parameter with null hypothesis $H_0$ given by $\btheta^*=\btheta_0$ and alternative hypothesis $H_1$ given by $\btheta^*\neq\btheta_0$, and report the $p$-value. We evaluate the test over a grid of candidate values of $\btheta_0$, where the distance between $\btheta_0$ and the true parameter $\btheta^*$ quantifies the magnitude of departure from the null. For each $\btheta_0$, we conduct the Wald test described in Section A of the Supplementary Material \citep{mppi2026} over 1000 simulated datasets and calculate the $p$-value. For mean-inference testing, we use 200 equally spaced null values between $\btheta^*-0.1$ and $\btheta^*+0.1$ for the homogeneous and covariate-shift settings, and between $\btheta^*-0.3$ and $\btheta^*+0.3$ for the domain-shift setting\footnote{The wider range in the domain-shift setting is used only to display the intersections of the flatter p-value curves with the nominal level $\alpha=0.05$.}.

\subsection{Simulation Results}

Tables~\ref{simu:homo-ACP-mean}-\ref{simu:heter-ACP-mean} report the ACP and  $\text{Vol}$ for mean inference under the homogeneous and heterogeneous settings, respectively. In these mean-inference designs, MPPI attains the smallest $\text{Vol}$ among the competing methods while maintaining Type I error control, as reflected by ACP values close to the nominal coverage level. These empirical patterns align with the theoretical results established in Corollaries~\ref{cor:MPPI_coverage}, \ref{cor:plugin_oracle_det}, \ref{cor:MPPI_coverage_covariate_shift}, \ref{cor:plugin_oracle_det_covariate_shift}, \ref{cor:MPPI_coverage_domain_shift}, and \ref{cor:plugin_oracle_det_domain_shift}. In addition, the results illustrate the benefit of aggregating multiple sources in the simulated settings. In the covariate-shift setting, some single-source PPI methods produce noticeably larger confidence regions (e.g., PPI with source~3). In contrast, MPPI uses information across sources and yields smaller confidence regions than the classical estimator and the PPI-type baselines in these designs. Under domain shift, MPPI has the smallest $\text{Vol}$ in both DGPs, although the performance gaps among MPPI, PPI, PPI++, and EW are more modest than those under covariate shift.

Table~\ref{simu:mean-weight} reports the average weights selected by MPPI on the target data and the source datasets. Under homogeneous distributions, the MPPI weights are mainly proportional to the sample sizes. Under covariate shift, MPPI assigns more weight to sources that lead to tighter inference for the target estimand. Under domain shift, the selected weights concentrate almost entirely on the transformed source data, primarily Sources~1 and~2, with negligible weights on the target sample and Source~3. This pattern reflects the construction of the source-specific machine learning methods $f^{(s)}$. Sources~1 and~2 use coordinate-specific features that capture the main signal components in $g_{\mathrm M}$, whereas Source~3 uses only an intercept and thus contains little covariate-dependent predictive information.  Overall, these data-driven weighting patterns help explain the observed coverage and confidence-region volume reductions in these simulation settings.

\renewcommand{\arraystretch}{1.2}
\begin{table}[!ht]
    \centering
    \caption{The ACP and $\text{Vol}$ in mean inference under homogeneous distribution} 
    \begin{tabular}{lcccc}
    \toprule
    \multirow{3}{*}{Methods} 
        & \multicolumn{2}{c}{DGP-linear} 
        & \multicolumn{2}{c}{DGP-nonlinear} \\ 
    \cmidrule(lr){2-3} \cmidrule(lr){4-5}
        & ACP & $\text{Vol}$ & ACP & $\text{Vol}$ \\ 
    \midrule
        MPPI  & 0.9510 & \textbf{1.1604} & 0.9470 & \textbf{1.8573}  \\ 
       Classic & 0.9440 & 3.0835 & 0.9390 & 6.1454  \\ 
        PPI (Source 1)  & 0.9510 & 2.0051 & 0.9480 & 3.7716  \\ 
        PPI (Source 2)  & 0.9550 & 1.5216 & 0.9600 & 2.6669  \\ 
        PPI (Source 3)  & 0.9460 & 1.3523 & 0.9540 & 2.2781  \\ 
        PPI++ (Source 1)  & 0.9470 & 1.6942 & 0.9410 & 3.0475  \\ 
        PPI++ (Source 2)  & 0.9450 & 1.4172 & 0.9550 & 2.4297  \\
        PPI++ (Source 3)  & 0.9490 & 1.2982 & 0.9590 & 2.1645  \\ 
        EW  & 0.9500 & 1.2507 & 0.9520 & 2.0577  \\ 
    \bottomrule
    \end{tabular}
    \label{simu:homo-ACP-mean}
\end{table}

\renewcommand{\arraystretch}{1.2}
\begin{table}[!ht]
    \centering
    \caption{The ACP and $\text{Vol}$ for mean inference under heterogeneous distribution}
    \resizebox{\textwidth}{!}{
    \begin{tabular}{lcccccccc}
    \toprule
    \multirow{3}{*}{Methods} & \multicolumn{4}{c}{Covariate shift} & \multicolumn{4}{c}{Domain shift} \\ \cmidrule(lr){2-5} \cmidrule(lr){6-9} 
        & \multicolumn{2}{c}{DGP-linear} 
        & \multicolumn{2}{c}{DGP-nonlinear} & \multicolumn{2}{c}{DGP-linear} 
        & \multicolumn{2}{c}{DGP-nonlinear}  \\ 
    \cmidrule(lr){2-3} \cmidrule(lr){4-5}  \cmidrule(lr){6-7} \cmidrule(lr){8-9} 
        & ACP & $\text{Vol}$ & ACP & $\text{Vol}$ 
        & ACP & $\text{Vol}$ & ACP & $\text{Vol}$ \\ 
    \midrule
        MPPI & 0.9560 & \textbf{1.7258} & 0.9560 & \textbf{3.2054} & 0.9400 & \textbf{50.0368} & 0.9440 & \textbf{49.7057}  \\ 
        Classic & 0.9440 & 3.0835 & 0.9390 & 6.1454 & 0.9410 & 51.9094 & 0.9450 & 50.8437  \\ 
        PPI (Source 1) & 0.9860 & 2.6981 & 0.9880 & 5.5465 & 0.9410 & 51.0566 & 0.9480 & 50.2819  \\ 
        PPI (Source 2) & 0.9580 & 5.1895 & 0.9430 & 12.9124 & 0.9400 & 50.8490 & 0.9450 & 50.1764  \\ 
        PPI (Source 3) & 0.9370 & 26.0782 & 0.8970 & 73.4177 & 0.9410 & 51.9094 & 0.9450 & 50.8437  \\ 
        PPI++ (Source 1) & 0.9540 & 1.9812 & 0.9580 & 3.7628 & 0.9410 & 50.9335 & 0.9470 & 50.2350  \\ 
        PPI++ (Source 2) & 0.9420 & 2.6337 & 0.9430 & 5.2580 & 0.9400 & 50.7538 & 0.9450 & 50.1247  \\ 
        PPI++ (Source 3) & 0.9470 & 2.9955 & 0.9380 & 5.9696 & 0.9410 & 51.9094 & 0.9450 & 50.8437  \\ 
        EW & 0.9570 & 3.1556 & 0.9420 & 7.5709 & 0.9400 & 50.7518 & 0.9450 & 50.1503  \\ 
    \bottomrule
    \end{tabular}}
    \label{simu:heter-ACP-mean}
\end{table}

\renewcommand{\arraystretch}{1.2}
\begin{table}[!ht]
    \centering
    \caption{Weight allocation of MPPI in mean inference}
    \resizebox{\textwidth}{!}{
    \begin{tabular}{ccccccc}
    \toprule
    \multirow{3}{*}{Model } 
        & \multicolumn{2}{c}{\multirow{2}{*}{Homogeneous}}
        & \multicolumn{4}{c}{Heterogeneous} \\ 
     \cmidrule(lr){4-7} 
       &  &  & \multicolumn{2}{c}{Covariate shift} & \multicolumn{2}{c}{Domain shift}  \\ 
       \cmidrule(lr){2-3} \cmidrule(lr){4-5} \cmidrule(lr){6-7}
       & DGP-linear & DGP-nonlinear & DGP-linear & DGP-nonlinear & DGP-linear & DGP-nonlinear \\ 
    \midrule
    Target & 0.0773 & 0.0770 & 0.3414 & 0.3623 & 0.0000 & 0.0000  \\ 
    Source 1 & 0.1568 & 0.1556 & 0.4206 & 0.4171 & 0.4713 & 0.4744  \\ 
    Source 2 & 0.3075 & 0.3069 & 0.1902 & 0.1748 & 0.5287 & 0.5256  \\ 
    Source 3 & 0.4584 & 0.4604 & 0.0478 & 0.0458 & 0.0000 & 0.0000  \\ 
    \bottomrule
    \end{tabular}}
    \label{simu:mean-weight}
\end{table}

\begin{figure}[htbp]
  \centering
  \includegraphics[width=0.9\linewidth]{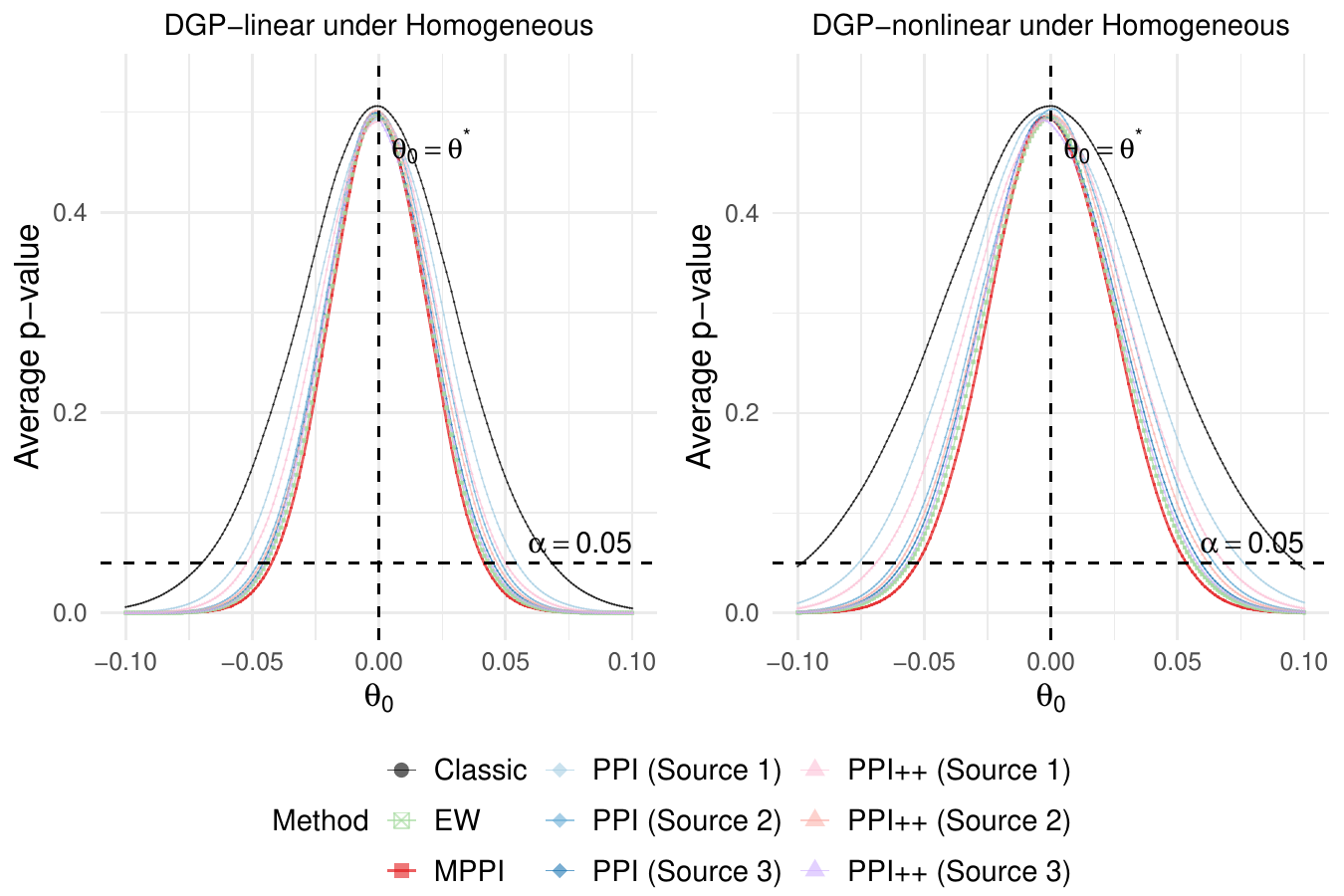}
  \caption{The relationship between $\btheta_0$ in the null hypothesis $H_0$ given by $\btheta^*=\btheta_0$ and the $p$-values for mean inference under homogeneous settings.}
  \label{fig:pvalue_meanhomo}
\end{figure}

\begin{figure}[htbp]
  \centering
    \includegraphics[width=0.9\linewidth]{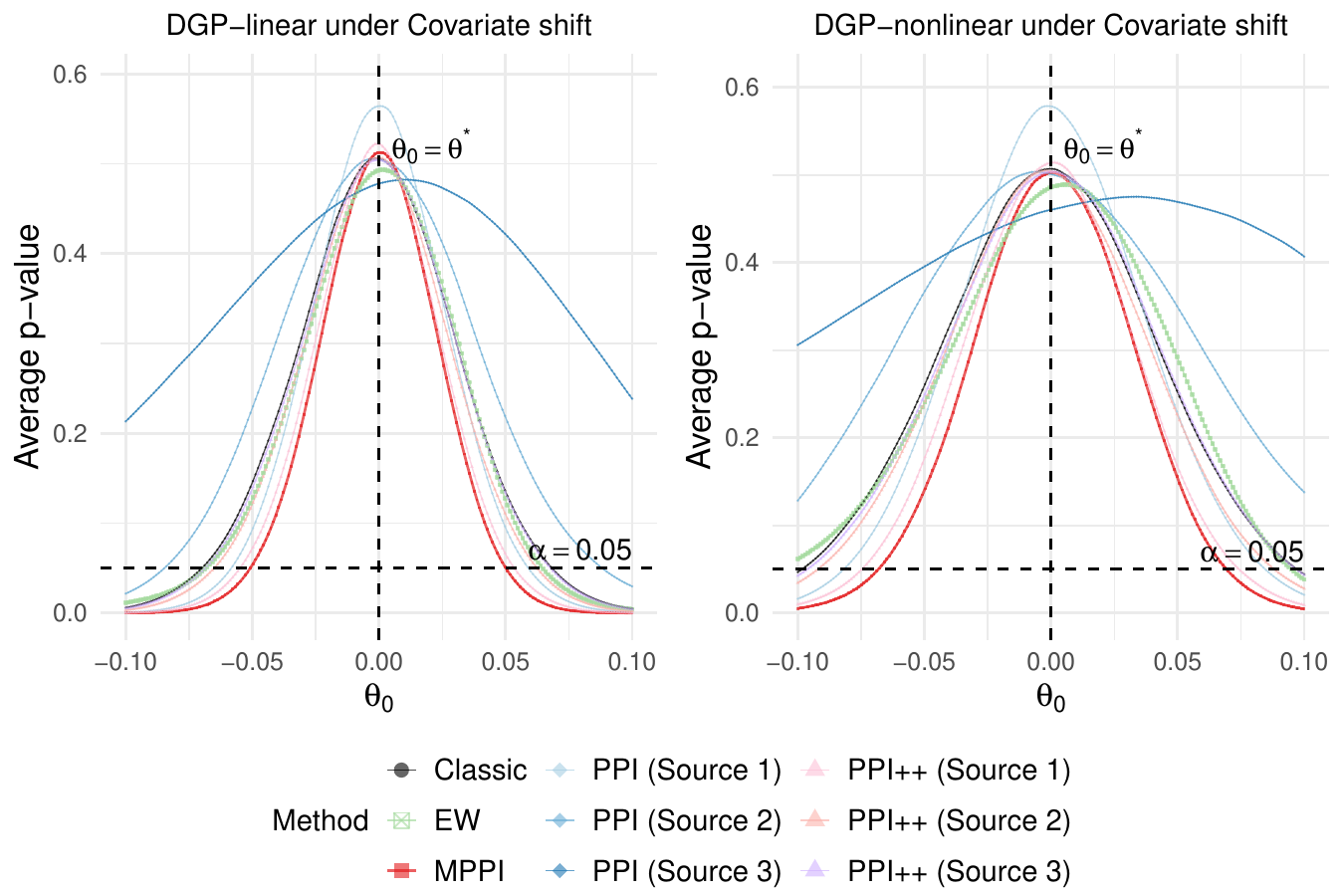}
  \caption{The relationship between $\btheta_0$ in the null hypothesis $H_0$ given by $\btheta^*=\btheta_0$ and the $p$-values for mean inference under covariate shift settings.}
  \label{fig:pvalue_mean_cov}
\end{figure}

\begin{figure}[htbp]
  \centering
    \includegraphics[width=0.9\linewidth]{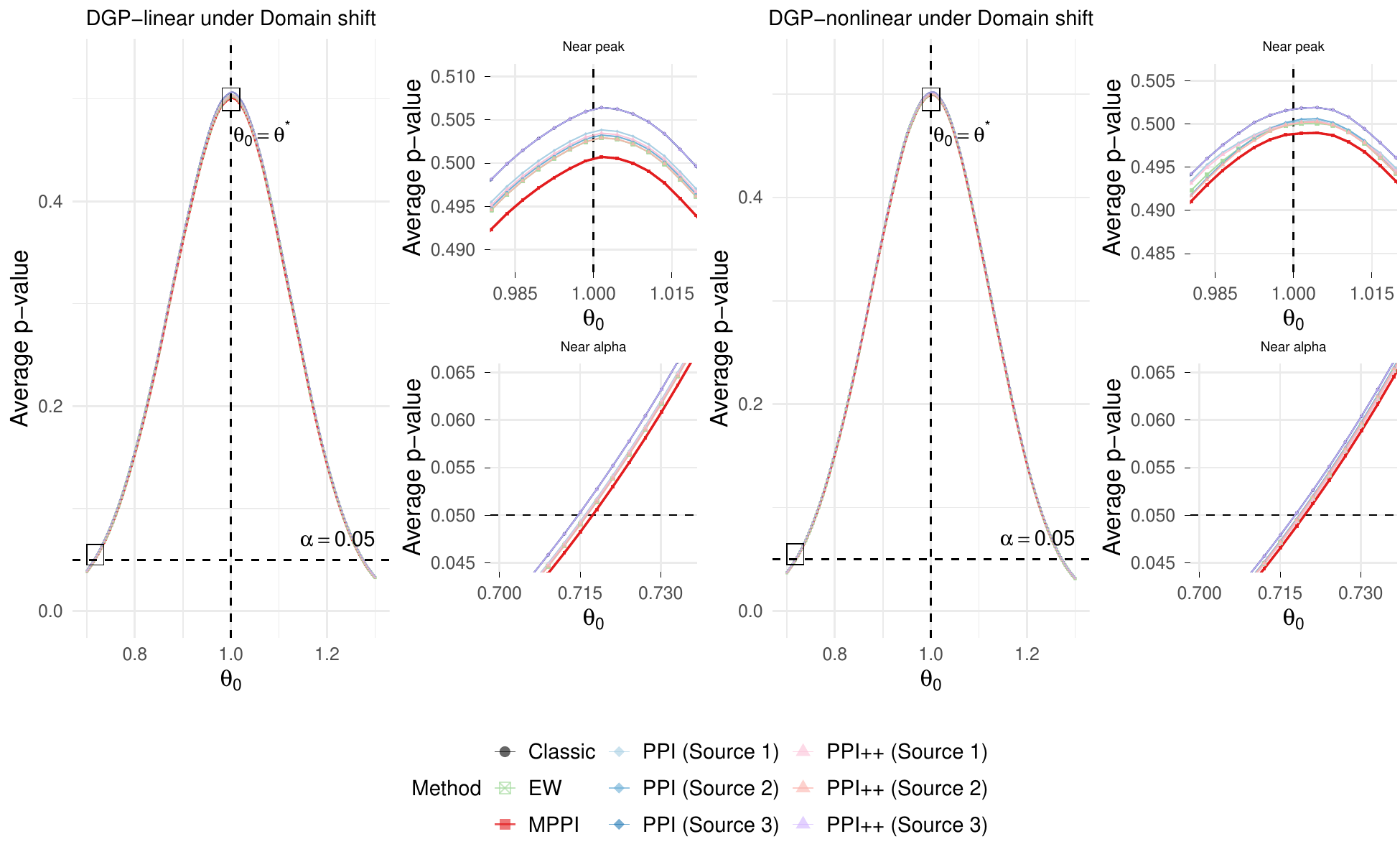}
  \caption{The relationship between $\btheta_0$ in the null hypothesis $H_0$ given by $\btheta^*=\btheta_0$ and the $p$-values for mean inference under the domain shift setting.}
  \label{fig:pvalue_mean_domain}
\end{figure}

The hypothesis testing results for mean inference are presented in Figures~\ref{fig:pvalue_meanhomo}, \ref{fig:pvalue_mean_cov}, and \ref{fig:pvalue_mean_domain}, with the corresponding results for linear regression inference deferred to Section E of the Supplementary Material \citep{mppi2026}. 
Overall, for our MPPI method, the $p$-value curves are bell-shaped and approximately symmetric around the true null value $\btheta_0=\btheta^*$, which is expected for a two-sided test. Under covariate shift, the $p$-value profiles of PPI and PPI++ using Source~3 exhibit noticeable asymmetry relative to $\btheta^*=0$.
In addition, the width of the curve reflects the sensitivity of the test to deviations from the null hypothesis. Under the homogeneous and covariate-shift settings, MPPI generally produces relatively narrow curves among the competing methods, indicating that its $p$-values decrease rapidly as $\btheta_0$ deviates from $\btheta^*$. 
Moreover, while controlling Type I error at the nominal level in these simulations, the $p$-values of MPPI generally fall below the rejection threshold at smaller deviations of $\btheta_0$ from $\btheta^*$ than those of other methods, suggesting earlier rejection of the null hypothesis in the displayed designs.
Under domain shift, the curves of all methods are nearly indistinguishable, suggesting that the methods have comparable testing behavior in this mean-inference setting.

\newpage 
\section{Inference for DXA-Measured High Body Fat Prevalence}
\label{sec:real_data}
In this section, we assess the empirical performance of the proposed method in a real-data application concerning the prevalence of high body fat. Let $Y$ denote total percent body fat measured by dual-energy X-ray absorptiometry (DXA), recorded on the percentage scale. For each sex-specific analysis, we define $Z_c=\mathbb{I}(Y>c)$ and consider the target parameter $\btheta_c^*=\mE(Z_c)=\mathrm{Pr}(Y>c)$. We focus on participants aged 40 to 59 years and use sex-specific high-adiposity thresholds, with $c=30$ for men and $c=40$ for women. These cutoffs can be interpreted as public-health-motivated high-adiposity thresholds linked to body mass index (BMI)-based obesity guidelines \citep{gallagher2000healthy, heo2012percentage}.  DXA provides a reliable measurement of body composition, but it is costly and is therefore typically available only for a relatively small sample size. This motivates the use of additional information from related source datasets to improve inference on the target parameter.

We use the 2017--2018 National Health and Nutrition Examination Survey (NHANES) whole-body DXA component as the labeled target dataset.\footnote{NHANES provides nationally collected health, demographic, and examination data for U.S. participants. The 2017--2018 whole-body DXA data are available at \url{https://wwwn.cdc.gov/Nchs/Data/Nhanes/Public/2017/DataFiles/DXX_J.htm}.} For the NHANES dataset, we focus on male and female participants aged 40 to 59 years and then exclude observations with missing total percent body fat or incomplete covariate information. The resulting labeled target samples contain $N_0=529$ male participants and $N_0=584$ female participants.
For the source datasets, we use three publicly available U.S. health survey datasets from the same 2017--2018 period that are unlabeled with respect to DXA-measured body fat, namely the National Health Interview Survey (NHIS),\footnote{NHIS is a national household health survey that collects information on health status, health care access, and health-related characteristics of the U.S. population. The 2017 and 2018 NHIS data releases are available at \url{https://archive.cdc.gov/www_cdc_gov/nchs/nhis/nhis_2017_data_release.htm} and \url{https://archive.cdc.gov/www_cdc_gov/nchs/nhis/nhis_2018_data_release.htm}.} the Behavioral Risk Factor Surveillance System (BRFSS),\footnote{BRFSS is a state-based telephone health survey system that collects data on health-related risk behaviors, chronic health conditions, and use of preventive services among U.S. adults. The 2017 and 2018 BRFSS annual data are available at \url{https://www.cdc.gov/brfss/annual_data/annual_2017.html} and \url{https://www.cdc.gov/brfss/annual_data/annual_2018.html}.} and the Health Information National Trends Survey (HINTS) 5 Cycles 1 and 2.\footnote{HINTS is a national survey program on health communication and cancer-related information. The public-use HINTS 5 Cycle 1 and Cycle 2 datasets are available at \url{https://hints.cancer.gov/data/download-data.aspx}.} These source datasets contain health-related covariates associated with body composition, but they do not contain DXA-measured total percent body fat. After applying the same age and sex restrictions and removing observations with missing covariates, the male source sample sizes are $7487$ for NHIS, $119843$ for BRFSS, and $886$ for HINTS, and the female source sample sizes are $8320$ for NHIS, $133665$ for BRFSS, and $1337$ for HINTS. To combine the target and source datasets, we use age and BMI as the common covariates in this application. Here, age is measured in years and BMI is measured in kg/m$^2$.

For the pretrained machine learning methods $f^{(s)}$, we use three published body-fat equations that depend only on variables available in all target and source datasets. Source 1 (NHIS) is paired with the Deurenberg equation \citep{deurenberg1991body}, Source 2 (BRFSS) is paired with the CUN-BAE equation \citep{gomezambrosi2012clinical}, and Source 3 (HINTS) is paired with the Gallagher equation \citep{gallagher2000healthy}. The corresponding pseudo-labels are $\mathbb{I}(f^{(s)}(X)>c)$, with the same sex-specific threshold $c$ used in the target estimand. Since the target and source samples come from different U.S. health survey programs, their covariate distributions may differ. We therefore use the covariate-shift version of MPPI, where the density ratio estimators $\widehat{\mathrm{DR}}^{(s)}(\cdot)$ are obtained by the cross-fitted discriminative approach described in Section D.3 of the Supplementary Material \citep{mppi2026}. Because each analysis conditions on sex, the covariate-shift adjustment is based on age and BMI. We then construct $(1-\alpha)$ confidence intervals for $\btheta_c^*$ using MPPI and the competing approaches described in Section~\ref{sec:sim}, with $\alpha=0.05$. For this scalar prevalence target, $\widetilde{\mathrm{Vol}}$ denotes the estimated asymptotic variance component $\widehat\Sigma$, with $\widehat{\mathrm{se}}(\widehat\btheta)=\sqrt{\widehat\Sigma/N_0}$. 

\renewcommand{\arraystretch}{1.2}
\begin{table}[!ht]
    \centering
    \caption{Inference results for DXA-measured high body fat prevalence}
    \label{appdxa:result} 
    \begin{tabular}{lcccc}
    \toprule
    Method & $\widehat{\btheta}$ & $\widehat{\mathrm{se}}(\widehat{\btheta})$ 
    & $\widetilde{\mathrm{Vol}}$ & 95\% CI \\
    \midrule
    \multicolumn{5}{l}{$\btheta^*_{\mathrm{male},30}=\mathrm{Pr}(Y>30\mid \mathrm{male},\,40\leq \mathrm{age}\leq 59)$} \\
    \midrule
    MPPI & 0.3950 & 0.0186 & \textbf{0.1830} & [0.3585, 0.4314] \\
    Classic & 0.3989 & 0.0213 & 0.2402 & [0.3571, 0.4406] \\
    PPI (Source 1, NHIS) & 0.3846 & 0.0232 & 0.2840 & [0.3392, 0.4300] \\
    PPI (Source 2, BRFSS) & 0.3892 & 0.0223 & 0.2628 & [0.3455, 0.4329] \\
    PPI (Source 3, HINTS) & 0.4008 & 0.0249 & 0.3275 & [0.3520, 0.4496] \\
    PPI++ (Source 1, NHIS) & 0.3929 & 0.0192 & 0.1943 & [0.3554, 0.4305] \\
    PPI++ (Source 2, BRFSS) & 0.3945 & 0.0189 & 0.1891 & [0.3574, 0.4315] \\
    PPI++ (Source 3, HINTS) & 0.3995 & 0.0201 & 0.2133 & [0.3601, 0.4389] \\
    EW & 0.3934 & 0.0190 & 0.1906 & [0.3562, 0.4306] \\
    \midrule
    \multicolumn{5}{l}{$\btheta^*_{\mathrm{female},40}=\mathrm{Pr}(Y>40\mid \mathrm{female},\,40\leq \mathrm{age}\leq 59)$} \\
    \midrule
    MPPI & 0.5610 & 0.0169 & \textbf{0.1666} & [0.5279, 0.5941] \\
    Classic & 0.5599 & 0.0206 & 0.2468 & [0.5196, 0.6002] \\
    PPI (Source 1, NHIS) & 0.5662 & 0.0200 & 0.2332 & [0.5270, 0.6053] \\
    PPI (Source 2, BRFSS) & 0.5602 & 0.0195 & 0.2218 & [0.5220, 0.5984] \\
    PPI (Source 3, HINTS) & 0.5546 & 0.0243 & 0.3441 & [0.5070, 0.6022] \\
    PPI++ (Source 1, NHIS) & 0.5632 & 0.0172 & 0.1735 & [0.5294, 0.5970] \\
    PPI++ (Source 2, BRFSS) & 0.5601 & 0.0171 & 0.1708 & [0.5266, 0.5936] \\
    PPI++ (Source 3, HINTS) & 0.5580 & 0.0188 & 0.2058 & [0.5213, 0.5948] \\
    EW & 0.5602 & 0.0174 & 0.1776 & [0.5260, 0.5944] \\
    \bottomrule
    \end{tabular}
\end{table}

The results are summarized in Table~\ref{appdxa:result}. For men, the classical target-only estimator gives $\widehat{\btheta}=0.3989$ with $\widetilde{\mathrm{Vol}}=0.2402$. MPPI gives a similar estimate, $\widehat{\btheta}=0.3950$, but reduces $\widetilde{\mathrm{Vol}}$ to 0.1830. For women, the classical estimator gives $\widehat{\btheta}=0.5599$ with $\widetilde{\mathrm{Vol}}=0.2468$, while MPPI gives $\widehat{\btheta}=0.5610$ and reduces $\widetilde{\mathrm{Vol}}$ to 0.1666. The estimated MPPI weights are $(0.4467,0.0917,0.3007,0.1608)$ for the male analysis and $(0.4128,0.2127,0.3197,0.0548)$ for the female analysis, where the entries correspond to the target sample, Source 1, Source 2, and Source 3. Thus, MPPI preserves the substantive conclusions from the target-only estimator while improving precision by combining the labeled DXA sample with informative unlabeled auxiliary data. Substantively, the estimates suggest that about two in five men and more than one half of women aged 40 to 59 years in this analytic target sample exceed the corresponding sex-specific high-adiposity threshold. Because survey design weights are not incorporated in this analysis, the results should be interpreted as evidence for the analytic target distribution rather than as official national prevalence estimates.

Additionally, we report the sensitivity analyses of other thresholds for male and female in Section G of the Supplementary Material \citep{mppi2026}. There, we vary the thresholds to $c=28$ and $c=32$ for men and to $c=38$ and $c=42$ for women, and MPPI continues to produce estimates close to the target-only estimator with smaller estimated variance.

\section{Discussion}\label{sec:conc}
This paper develops a general MPPI framework for conducting statistically valid inference by leveraging machine-learning methods to construct multiple auxiliary datasets.  We provide a unified theoretical analysis establishing asymptotic normality and valid confidence regions  under homogeneous and heterogeneous distributions. 
Moreover, we characterize interpretable sufficient conditions under which MPPI and PPI can achieve asymptotic efficiency gains relative to the classical estimator based solely on the target data, thereby clarifying when auxiliary information are beneficial for obtaining tighter confidence region.

\subsection{Scope of PPI-Type Methods}\label{sec:scope}
We close by clarifying the scope of PPI-type methods. The main point is that pseudo-labels can improve efficiency only through the part of the true-label directional score that is predictable from covariates. Based on the notation and assumptions in the homogeneous setting in Section~\ref{sec:homogeneous}, for any direction $\bu$, decompose the true-label directional score as
\begin{align*}
G_{\btheta^*}(\bu)=\mE(G_{\btheta^*}(\bu)\mid X^{(0)})+\Bigl(G_{\btheta^*}(\bu)-\mE(G_{\btheta^*}(\bu)\mid X^{(0)})\Bigr)=\mu_{\btheta^*}(\bu\mid X^{(0)})+\eta_{\btheta^*}(\bu),
\end{align*}
where $\mu_{\btheta^*}(\bu\mid X^{(0)})=\mE(G_{\btheta^*}(\bu)\mid X^{(0)})$ is the covariate-explainable component of the directional score, and $\eta_{\btheta^*}(\bu)$ is the residual variation induced by the conditional randomness of $Y^{(0)}$ given $X^{(0)}$. Since $G_{\btheta^*}^{(s)}(\bu)$ is generated from $f^{(s)}(X^{(0)})$, it is a function of $X^{(0)}$ only. Thus, it can align with the component $\mu_{\btheta^*}(\bu\mid X^{(0)})$, but it cannot capture the residual component $\eta_{\btheta^*}(\bu)$ because $\mE(\eta_{\btheta^*}(\bu)\mid X^{(0)})=0$. More generally, for any pseudo-label directional score that is a function of $X^{(0)}$, its covariance with $G_{\btheta^*}(\bu)$ is its covariance with $\mu_{\btheta^*}(\bu\mid X^{(0)})$. In particular,
$\cov(G_{\btheta^*}(\bu),G_{\btheta^*}^{(s)}(\bu))=\cov(\mu_{\btheta^*}(\bu\mid X^{(0)}),G_{\btheta^*}^{(s)}(\bu))$ and $\cov(G_{\btheta^*}(\bu),G_{\btheta^*}(\bu;\bw))=\cov(\mu_{\btheta^*}(\bu\mid X^{(0)}),G_{\btheta^*}(\bu;\bw))$.
Therefore, the covariance term in the projection coefficient is driven by how well the pseudo-label directional score, or the weighted pseudo-label directional score in MPPI, aligns with the covariate-explainable score component. If the true-label directional score is mostly residual conditional noise, then there is little covariate-driven score structure for pseudo-labels to exploit, and PPI-type methods should provide limited efficiency gain over the classical estimator.

The following example illustrates why this scope restriction matters.
\begin{example}\label{remark1}
Consider the data-generating process $Y=g(X)+\epsilon$ with $\mE(\epsilon\mid X)=0$. Suppose the inferential target is the ordinary least squares (OLS) estimand under the linear working model $X^\top\btheta$ with squared loss $\ell_{\btheta}(X,Y)=\tfrac12(Y-X^\top\btheta)^2$. Let $\btheta^*=\argmin_{\btheta}\mE((Y-X^\top\btheta)^2)$, which may be the pseudo-true parameter. Then
$G_{\btheta^*}(\bu)=-(\bu^\top X)(Y-X^\top\btheta^*)$ and $G_{\btheta^*}^{(s)}(\bu)=-(\bu^\top X)(f^{(s)}(X)-X^\top\btheta^*)$.
If the linear working model is correctly specified, namely $g(X)=X^\top\btheta^*$, then $G_{\btheta^*}(\bu)=-(\bu^\top X)\epsilon$ and $\mE(G_{\btheta^*}(\bu)\mid X)=0$. Hence the covariate-explainable score component is degenerate. Any pseudo-label directional score generated from $f^{(s)}(X)$ is a function of $X$ only and has zero covariance with $G_{\btheta^*}(\bu)$. Consequently, the sufficient condition \eqref{eq:rho-threshold} cannot hold for a nondegenerate pseudo-label directional score. In contrast, if the linear working model is misspecified and $r(X):=g(X)-X^\top\btheta^*$ is nonconstant, then $Y-X^\top\btheta^*=r(X)+\epsilon$ and $\mE(G_{\btheta^*}(\bu)\mid X)=-(\bu^\top X)r(X)$. In this case, the true-label directional score contains a nontrivial covariate-explainable component. When $f^{(s)}(X)$ accurately approximates the conditional mean $g(X)$, the pseudo-label directional score can be well aligned with this component, making the condition \eqref{eq:rho-threshold} more likely to hold.
\end{example}
Example~\ref{remark1} shows that PPI-type improvements depend on score-relevant covariate structure, not only on the availability of auxiliary pseudo-labeled data. When the loss and working model already remove the conditional structure of $Y^{(0)}$ given $X^{(0)}$ at $\btheta=\btheta^*$, the remaining score variation is irreducible from the perspective of pseudo-labels based only on $X^{(0)}$. When the working model leaves a nondegenerate covariate-explainable component in the score, a sufficiently accurate pseudo-labeling method may reduce that component and yield efficiency gains. In this sense, PPI-type methods are most useful when the pseudo-label directional scores capture score-relevant information that is explainable by covariates.

\subsection{Future Work} 
Several directions remain for future research. 
First, it is of interest to extend the MPPI framework to settings where auxiliary data are generated by modern generative models, potentially including both covariates and responses.  Second, it would be valuable to extend MPPI to downstream decision-making objectives, in order to support reliable decision rules.

\bibliography{bibliography.bib}

@article{Anastasios2023science,
author = {Angelopoulos, Anastasios N. and Bates, Stephen and Fannjiang, Clara and Jordan, Michael I. and Zrnic, Tijana},
title = {Prediction-powered inference},
journal = {Science},
volume = {382},
number = {6671},
pages = {669--674},
year = {2023}
}

@inproceedings{huang2006NIPS,
 author = {Huang, Jiayuan and Gretton, Arthur and Borgwardt, Karsten and Sch\"{o}lkopf, Bernhard and Smola, Alex},
 booktitle = {Advances in Neural Information Processing Systems},
 editor = {B. Sch\"{o}lkopf and J. Platt and T. Hoffman},
 pages = {601--608},
 publisher = {MIT Press},
 title = {Correcting Sample Selection Bias by Unlabeled Data},
 volume = {19},
 year = {2006}
}

@article{Sugiyama2008AISM,
  title={Direct importance estimation for covariate shift adaptation},
  author={Sugiyama, Masashi and Suzuki, Taiji and Nakajima, Shinichi and Kashima, Hisashi and Von B{\"u}nau, Paul and Kawanabe, Motoaki},
  journal={Annals of the Institute of Statistical Mathematics},
  volume={60},
  pages={699--746},
  year={2008},
  publisher={Springer}
}

@article{Bickel2009JMLR,
  title={Discriminative learning under covariate shift},
  author={Bickel, Steffen and Br{\"u}ckner, Michael and Scheffer, Tobias},
  journal={Journal of Machine Learning Research},
  volume={10},
  number={9},
  year={2009},
  pages={2137--2155}
}

@misc{angelopoulos2024PPIplus,
  title         = {{PPI++}: Efficient Prediction-Powered Inference},
  author        = {Angelopoulos, Anastasios N. and Duchi, John C. and Zrnic, Tijana},
  year          = {2024},
  eprint        = {2311.01453},
  archivePrefix = {arXiv},
  primaryclass  = {stat.ML},
  url           = {https://arxiv.org/abs/2311.01453},
  note          = {arXiv:2311.01453}
}

@article{zrnic2024cppi, 
author = {Zrnic, Tijana and Cand{\`e}s, Emmanuel J.},
title = {Cross-prediction-powered inference},
journal = {Proceedings of the National Academy of Sciences},
volume = {121},
number = {15},
pages = {e2322083121},
year = {2024}
}

@incollection{sun2017coral,
  author    = {Sun, Baochen and Feng, Jiashi and Saenko, Kate},
  title     = {Correlation Alignment for Unsupervised Domain Adaptation},
  booktitle = {Domain Adaptation in Computer Vision Applications},
  pages     = {153--171},
  year      = {2017},
  publisher = {Springer International Publishing},
  doi       = {10.1007/978-3-319-58347-1_8}
}

@inproceedings{tzeng2017adda,
  title     = {Adversarial Discriminative Domain Adaptation},
  author    = {Tzeng, Eric and Hoffman, Judy and Saenko, Kate and Darrell, Trevor},
  booktitle = {Proceedings of the IEEE Conference on Computer Vision and Pattern Recognition},
  year      = {2017},
  pages     = {7167--7176}
}

@inproceedings{seguy2017large,
  author    = {Seguy, Vivien and Damodaran, Bharath Bhushan and Flamary, Remi and Courty, Nicolas and Rolet, Antoine and Blondel, Mathieu},
  title     = {Large-Scale Optimal Transport and Mapping Estimation},
  booktitle = {International Conference on Learning Representations},
  year      = {2018},
  address   = {Vancouver, Canada},
  url       = {https://openreview.net/forum?id=B1zlp1bRW}
}

@inproceedings{makkuva2020optimal,
  title     = {Optimal Transport Mapping via Input Convex Neural Networks},
  author    = {Makkuva, Ashok and Taghvaei, Amirhossein and Oh, Sewoong and Lee, Jason D.},
  booktitle = {Proceedings of the 37th International Conference on Machine Learning},
  series    = {Proceedings of Machine Learning Research},
  volume    = {119},
  pages     = {6672--6681},
  year      = {2020},
  publisher = {PMLR}
}

@article{divol2022unbalanced,
author = {Vincent Divol and Jonathan Niles-Weed and Aram-Alexandre Pooladian},
title = {Optimal transport map estimation in general function spaces},
volume = {53},
journal = {The Annals of Statistics},
number = {3},
publisher = {Institute of Mathematical Statistics},
pages = {963--988},
year = {2025}
}

@inproceedings{deb2021wasserstein,
  title     = {Rates of Estimation of Optimal Transport Maps using Plug-in Estimators via Barycentric Projections},
  author    = {Deb, Nabarun and Ghosal, Promit and Sen, Bodhisattva},
  booktitle = {Advances in Neural Information Processing Systems},
  volume    = {34},
  pages     = {29736--29753},
  year      = {2021}
}

@inproceedings{Tibshirani2019NIPS,
  author    = {Tibshirani, Ryan J. and Foygel Barber, Rina and Cand{\`e}s, Emmanuel and Ramdas, Aaditya},
  title     = {Conformal Prediction Under Covariate Shift},
  booktitle = {Advances in Neural Information Processing Systems},
  editor    = {Wallach, H. and Larochelle, H. and Beygelzimer, A. and d'Alch{\'e}-Buc, F. and Fox, E. and Garnett, R.},
  volume    = {32},
  pages     = {2530--2540},
  publisher = {Curran Associates, Inc.},
  year      = {2019}
}

@article{yuan2025optimal,
  title={Optimal Transport based Cross-Domain Integration for Heterogeneous Data},
  author={Yuan, Yubai and Zhang, Yijiao and Shahbaba, Babak and Fortin, Norbert and Cooper, Keiland and Nie, Qing and Qu, Annie},
  journal={Journal of the American Statistical Association},
  volume={120},
  number={551},
  pages={1449--1462},
  year={2025},
  publisher={Taylor \& Francis}
}

@article{qin2025distribution,
  title={Distribution-free prediction intervals under covariate shift, with an application to causal inference},
  author={Qin, Jing and Liu, Yukun and Li, Moming and Huang, Chiung-Yu},
  journal={Journal of the American Statistical Association},
  volume={120},
  number={549},
  pages={559--571},
  year={2025},
  publisher={Taylor \& Francis}
}

@misc{jin2025model,
  title={Model-free selective inference under covariate shift via weighted conformal $p$-values},
  author={Jin, Ying and Cand{\`e}s, Emmanuel J},
  year={2023},
  eprint={2307.09291},
  archivePrefix={arXiv},
  primaryClass={stat.ME},
  url={https://arxiv.org/abs/2307.09291},
  note={arXiv:2307.09291}
}

@article{ge2024optimal,
  title={Optimal aggregation of prediction intervals under unsupervised domain shift},
  author={Ge, Jiawei and Mukherjee, Debarghya and Fan, Jianqing},
  journal={Advances in Neural Information Processing Systems},
  volume={37},
  pages={73605--73637},
  year={2024}
}

@article{sugiyama2007covariate,
  title={Covariate shift adaptation by importance weighted cross validation},
  author={Sugiyama, Masashi and Krauledat, Matthias and M{\"u}ller, Klaus-Robert},
  journal={Journal of Machine Learning Research},
  volume={8},
  number={5},
  pages={985-1005},
  year={2007}
}

@incollection{gretton2009covariate,
  title={Covariate shift by kernel mean matching},
  author={Gretton, Arthur and Smola, Alex and Huang, Jiayuan and Schmittfull, Marcel and Borgwardt, Karsten M. and Sch{\"o}lkopf, Bernhard},
  booktitle={Dataset Shift in Machine Learning},
  editor={Qui{\~n}onero-Candela, Joaquin and Sugiyama, Masashi and Schwaighofer, Anton and Lawrence, Neil D.},
  pages={131--160},
  publisher={MIT Press},
  year={2009}
}

@book{chapelle2006ssl,
    editor = {Chapelle, Olivier and Sch{\"o}lkopf, Bernhard and Zien, Alexander},
    title = {Semi-Supervised Learning},
    publisher = {The MIT Press},
    year = {2006},
    month = {09},
    isbn = {9780262255899},
    doi = {10.7551/mitpress/9780262033589.001.0001},
    url = {https://doi.org/10.7551/mitpress/9780262033589.001.0001},
}

@article{Jumper2021AlphaFold,
  title   = {Highly accurate protein structure prediction with {AlphaFold}},
  author  = {Jumper, John and Evans, Richard and Pritzel, Alexander and Green, Tim and Figurnov, Michael and Ronneberger, Olaf and Tunyasuvunakool, Kathryn and Bates, Russ and {\v{Z}}{\'i}dek, Augustin and Potapenko, Anna and Bridgland, Alex and Meyer, Clemens and Kohl, Simon A. A. and Ballard, Andrew J. and Cowie, Andrew and Romera-Paredes, Bernardino and Nikolov, Stanislav and Jain, Rishub and Adler, Jonas and Back, Trevor and Petersen, Stig and Reiman, David and Clancy, Ellen and Zielinski, Michal and Steinegger, Martin and Pacholska, Michalina and Berghammer, Tamas and Bodenstein, Sebastian and Silver, David and Vinyals, Oriol and Senior, Andrew W. and Kavukcuoglu, Koray and Kohli, Pushmeet and Hassabis, Demis},
  journal = {Nature},
  year    = {2021},
  volume  = {596},
  number  = {7873},
  pages   = {583--589},
  doi     = {10.1038/s41586-021-03819-2}
}

@article{Jean2016PovertySatellite,
  title   = {Combining satellite imagery and machine learning to predict poverty},
  author  = {Jean, Neal and Burke, Marshall and Xie, Michael and Davis, W. Matthew Alampay and Lobell, David B. and Ermon, Stefano},
  journal = {Science},
  year    = {2016},
  volume  = {353},
  number  = {6301},
  pages   = {790--794},
  doi     = {10.1126/science.aaf7894}
}

@article{wan2010least,
  title={Least squares model averaging by {Mallows} criterion},
  author={Wan, Alan TK and Zhang, Xinyu and Zou, Guohua},
  journal={Journal of Econometrics},
  volume={156},
  number={2},
  pages={277--283},
  year={2010},
  publisher={Elsevier}
}

@article{tunyasuvunakool2021highly,
  title={Highly accurate protein structure prediction for the human proteome},
  author={Tunyasuvunakool, Kathryn and Adler, Jonas and Wu, Zachary and Green, Tim and Zielinski, Michal and {\v{Z}}{\'\i}dek, Augustin and Bridgland, Alex and Cowie, Andrew and Meyer, Clemens and Laydon, Agata and others},
  journal={Nature},
  volume={596},
  number={7873},
  pages={590--596},
  year={2021},
  publisher={Nature Publishing Group UK London}
}

@article{lin2023evolutionary,
  title={Evolutionary-scale prediction of atomic-level protein structure with a language model},
  author={Lin, Zeming and Akin, Halil and Rao, Roshan and Hie, Brian and Zhu, Zhongkai and Lu, Wenting and Smetanin, Nikita and Verkuil, Robert and Kabeli, Ori and Shmueli, Yaniv and others},
  journal={Science},
  volume={379},
  number={6637},
  pages={1123--1130},
  year={2023},
  publisher={American Association for the Advancement of Science}
}

@article{zheng2023chatgpt,
  title={{ChatGPT} chemistry assistant for text mining and the prediction of {MOF} synthesis},
  author={Zheng, Zhiling and Zhang, Oufan and Borgs, Christian and Chayes, Jennifer T and Yaghi, Omar M},
  journal={Journal of the American Chemical Society},
  volume={145},
  number={32},
  pages={18048--18062},
  year={2023},
  publisher={ACS Publications}
}

@article{bolte2014pam,
  title={Proximal alternating linearized minimization for nonconvex and nonsmooth problems},
  author={Bolte, J{\'e}r{\^o}me and Sabach, Shoham and Teboulle, Marc},
  journal={Mathematical Programming},
  volume={146},
  number={1},
  pages={459--494},
  year={2014},
  publisher={Springer}
}

@article{attouch2010lpam,
  title={Proximal alternating minimization and projection methods for nonconvex problems: An approach based on the Kurdyka-{\L}ojasiewicz inequality},
  author={Attouch, H{\'e}dy and Bolte, J{\'e}r{\^o}me and Redont, Patrick and Soubeyran, Antoine},
  journal={Mathematics of Operations Research},
  volume={35},
  number={2},
  pages={438--457},
  year={2010},
  publisher={INFORMS}
}

@article{Zhang2016jasa,
author = {Xinyu Zhang and Dalei Yu and Guohua Zou and Hua Liang},
title = {Optimal Model Averaging Estimation for Generalized Linear Models and Generalized Linear Mixed-Effects Models},
journal = {Journal of the American Statistical Association},
volume = {111},
number = {516},
pages = {1775--1790},
year = {2016},
publisher = {Taylor \& Francis}
}

@article{deurenberg1991body,
  title={Body mass index as a measure of body fatness: Age- and sex-specific prediction formulas},
  author={Deurenberg, Paul and Weststrate, Jan A and Seidell, Jaap C},
  journal={British Journal of Nutrition},
  volume={65},
  number={2},
  pages={105--114},
  year={1991},
  publisher={Cambridge University Press}
}

@article{gomezambrosi2012clinical,
  title={Clinical usefulness of a new equation for estimating body fat},
  author={G{\'o}mez-Ambrosi, Javier and Silva, Camilo and Catal{\'a}n, Victoria and Rodr{\'\i}guez, Amaia and Galofr{\'e}, Juan Carlos and Escalada, Javier and Valent{\'\i}, Victor and Rotellar, Fernando and Romero, Sonia and Ram{\'\i}rez, Beatriz and others},
  journal={Diabetes Care},
  volume={35},
  number={2},
  pages={383--388},
  year={2012},
  publisher={American Diabetes Association}
}

@article{gallagher2000healthy,
author = {Dympna Gallagher and Steven B Heymsfield and Moonseong Heo and Susan A Jebb and Peter R Murgatroyd and Yoichi Sakamoto},
title = {Healthy percentage body fat ranges: An approach for developing guidelines based on body mass index},
journal = {The American Journal of Clinical Nutrition},
volume = {72},
number = {3},
pages = {694--701},
year = {2000}
}

@article{heo2012percentage,
  title={Percentage of body fat cutoffs by sex, age, and race-ethnicity in the {US} adult population from {NHANES} 1999--2004},
  author={Heo, Moonseong and Faith, Myles S and Pietrobelli, Angelo and Heymsfield, Steven B},
  journal={The American Journal of Clinical Nutrition},
  volume={95},
  number={3},
  pages={594--602},
  year={2012},
  publisher={Oxford University Press}
}

@misc{kato2024double,
      title={Double Debiased Covariate Shift Adaptation Robust to Density-Ratio Estimation}, 
      author={Masahiro Kato and Kota Matsui and Ryo Inokuchi},
      year={2024},
      eprint={2310.16638},
      archivePrefix={arXiv},
      primaryClass={stat.ME},
      url={https://arxiv.org/abs/2310.16638}, 
}

@misc{shan2026sadasafeadaptiveaggregation,
      title={{SADA}: Safe and adaptive aggregation of multiple black-box predictions in semi-supervised learning}, 
      author={Jiawei Shan and Zhifeng Chen and Yiming Dong and Yazhen Wang and Jiwei Zhao},
      year={2026},
      eprint={2509.21707},
      archivePrefix={arXiv},
      primaryClass={stat.ML},
      url={https://arxiv.org/abs/2509.21707}, 
}

@misc{mppi2026,
  author = {Li, Wenhui and Jiang, Fen and Zhang, Xinyu},
  year   = {2026},
  title  = {Supplement to ``{M}ulti-{S}ource {P}rediction-{P}owered {I}nference''}
}

\end{document}